\documentclass[pdftex,twocolumn,epjc3]{svjour3}          

\RequirePackage[T1]{fontenc}

\smartqed  
\usepackage{comment}
\usepackage{color}
\usepackage{parskip}
\usepackage{appendix}
\usepackage{amssymb}
\usepackage{booktabs}
\usepackage{comment}
\usepackage{graphicx}
\usepackage{caption}
\usepackage{subcaption}
\usepackage{mathrsfs}
\usepackage{tabularx}
\usepackage{arydshln}
\usepackage{amsmath}
\usepackage{multirow}
\usepackage{multicol}
\usepackage{adjustbox}

\RequirePackage{graphicx}
\RequirePackage{mathptmx}      
\RequirePackage{flushend}
\RequirePackage[numbers,sort&compress]{natbib}
\RequirePackage[colorlinks,citecolor=blue,urlcolor=blue,linkcolor=blue]{hyperref}

\journalname{Eur. Phys. J. C}
\makeatletter
\renewcommand*\env@matrix[1][*\c@MaxMatrixCols c]{%
  \hskip -\arraycolsep
  \let\@ifnextchar\new@ifnextchar
  \array{#1}}
\makeatother

\begin{document}

\title{Uncertainty components in profile likelihood fits
}


\author{Andr\'{e}s Pinto\thanksref{e1,addr1, addr2}
        \and
        Zhibo Wu\thanksref{e2,addr1,addr3} 
        \and
        Fabrice Balli\thanksref{e3,addr1}
        \and
        Nicolas Berger\thanksref{e4,addr4}
        \and
        Maarten Boonekamp\thanksref{e5,addr1, addr2}
        \and
        \'{E}milien Chapon\thanksref{e6,addr1}
        \and
        Tatsuo Kawamoto\thanksref{e7,addr3}
        \and
        Bogdan Malaescu\thanksref{e8,addr5}
}

\thankstext{e1}{e-mail: andres.eloy.pinto.pinoargote@cern.ch}
\thankstext{e2}{e-mail: zhibo.wu@cern.ch}
\thankstext{e3}{e-mail: fabrice.balli@cern.ch}
\thankstext{e4}{e-mail: nicolas.berger@cern.ch}
\thankstext{e5}{e-mail: Maarten.Boonekamp@cern.ch}
\thankstext{e6}{e-mail: emilien.chapon@cern.ch}
\thankstext{e7}{e-mail: tatsuo.kawamoto@cern.ch}
\thankstext{e8}{e-mail: Bogdan.Malaescu@cern.ch}

\institute{Universit\'{e} Paris-Saclay, CEA, D\'{e}partement de Physique des Particules, 91191, Gif-sur-Yvette, France. \label{addr1}
           \and
           Institut f\"{u}r Physik, Universit\"{a}t Mainz,  Staudingerweg 7, 55128 Mainz, Germany. \label{addr2}
           \and
           Department of Modern Physics, University of Science and Technology of China, Hefei, Anhui, China\label{addr3}
           \and
           LAPP, Univ. Savoie Mont Blanc, CNRS/IN2P3, Annecy\label{addr4}
           \and
           LPNHE, Sorbonne Université, Université Paris Cité, CNRS/IN2P3, Paris, France, 75252\label{addr5}
}

\date{Received: date / Accepted: date}

\maketitle

\begin{abstract}
When a measurement of a physical quantity is reported, the total uncertainty is usually decomposed into statistical and systematic uncertainties. This decomposition is not only useful to understand the contributions to the total uncertainty, but also required to propagate these contributions in subsequent analyses, such as combinations or interpretation fits including results from other measurements or experiments. In profile-likelihood fits, widely applied in high-energy physics analyses, contributions of systematic uncertainties are routinely quantified using ``impacts'', which are not adequate for such applications. We discuss the difference between impacts and actual uncertainty components, and establish methods to determine the latter in a wide range of statistical models.
\end{abstract}
\tableofcontents
\section{Introduction}
\label{sec:intro}

Measurement results are usually reported not only quoting the total uncertainty on the measured values, but also their breakdown into uncertainty components -- usually the statistical uncertainty, and one or more components of systematic uncertainty. A consistent propagation of uncertainties is of upmost importance for global analyses of measurement data, such as, for example, the determination of the anomalous magnetic moment of the muon~\cite{Davier:2010nc}, of the parton distribution functions of the proton~\cite{Amoroso:2022eow}, measurements of $Z$-boson properties at LEP1~\cite{ALEPH:2005ab}, and of the top-quark mass~\cite{CMS:2022kqg} and Higgs-boson properties~\cite{ATLAS:2016neq} at the LHC. In high-energy physics experiments, different techniques are used for obtaining this decomposition, depending on (but not fundamentally related to) the test statistic used to obtain the results.

The simplest statistical method consists in comparing a measured quantity or distribution to a model, parameterised only in terms of the physical constants to be determined. Auxiliary parameters (detector calibrations, theoretical predictions, etc) on which the model depends are fixed to their best estimates. The measured values of the physical constants result from the maximization of the corresponding likelihood. The curvature of the likelihood around its maximum is determined only by the expected fluctuations of the data, and yields the statistical uncertainty of the measurement\footnote{The curvature of the likelihood around its maximum only provides a lower bound on the standard deviation of the estimator in the general case (Cram\'{e}r--Rao inequality). Much of the discussion in this paper will be about the maximum likelihood estimator, which is asymptotically efficient, i.e. for which the equality is reached.}. 
Systematic uncertainties are obtained by repeating the procedure with varied models, obtained from the variation of the auxiliary parameters within their uncertainty, one parameter at a time~\cite{vanDyk:2023tqz}. Each variation represents a given source of uncertainty. The corresponding uncertainties in the final result are usually uncorrelated by construction, and are summed in quadrature to obtain the total measurement uncertainty.

When using this method, different measurements of the same physical constants can be readily combined. When all uncertainties are Gaussian, the Best Linear Unbiased Estimate (BLUE)~\cite{Lyons:1988rp,Valassi:2003mu} results from the analytical maximization of the joint likelihood of the input measurements, and unambiguously propagates the statistical and systematic uncertainties in the input measurements to the combined result. 

An improved statistical method consists in parameterising the model in terms of both the physical constants and the sources of uncertainty~\cite{Rolke:2004mj,ALEPH:2013dgf}, and has become a standard in LHC analysis. In this case, the maximum of the likelihood represents a global optimum for the physical constants and the uncertainty parameters, and determines their best values simultaneously. The curvature of the likelihood at its maximum reflects the fluctuations of the data and of the other sources of uncertainty, therefore giving the total uncertainty in the final result. 

The determination of the statistical and systematic uncertainty components in numerical profile-likelihood fits is the subject of the present note. Current practice universally employs so-called impacts~\cite{babarstat,fisher_1919,fisher_1921}, obtained as the quadratic difference between the total uncertainties of fits including or excluding a given source of uncertainty. While impacts quantify the increase in total uncertainty when including a new systematic source in a measurement, they cannot be interpreted as the contribution of this source to the total uncertainty in the complete measurement. Impacts do not add up to the total uncertainty, and do not match usual uncertainty decomposition formulas~\cite{Valassi:2003mu} even when they should, i.e. when all uncertainties are genuinely Gaussian.

These statements are illustrated with a simple example in Section~\ref{sec:simpleExample}. Sections~\ref{sec:unccov} and~\ref{sec:uncnp} summarize parameter estimation in the Gaussian approximation. Sources of uncertainty can be entirely encoded in the covariance matrix of the measurements (the "covariance representation"), or parameterised using nuisance parameters (the "nuisance parameter representation"). The equivalence between the approaches is recalled, and a detailed discussion of the fit uncertainties and correlations is provided. A consistent method for the decomposition of uncertainties in profile-likelihood ratio fits is introduced in Section~\ref{sec:UncDecomp_ShiftedObs}. The method is general as it results from a Taylor expansion of the likelihood, and a proof that it yields consistent results in the Gaussian regime is given. The different approaches are illustrated in Section~\ref{sec:examples} with examples based on the Higgs and $W$-boson mass measurements and combinations, which are usually dominated by systematic uncertainties and where the present discussion is of particular relevance. Concluding remarks are presented in Section~\ref{sec:conclusion}.

In the following, we understand the statistical uncertainty in its strict frequentist definition, i.e. the standard deviation of an estimator when the exact same experiment is repeated (with the same systematic uncertainties) on independent data samples of identical expected size. Similarly, a systematic uncertainty contribution should match the standard deviation of the estimator obtained under fluctuations of the corresponding source within its initial uncertainty. Measurements (physical parameters, cross sections or bins of a measured distribution) and the corresponding predictions will be denoted $\vec{m}$ and $\vec{t}$, respectively, and labeled using roman indices $i,j,k$. The predictions are functions of the physical constants to be determined, referred to as parameters of interest (POIs), denoted $\vec{\theta}$ and labeled $p,q$. Sources of uncertainty are denoted $\vec{a}$ and their associated nuisance parameters (NPs), $\vec{\alpha}$, are labeled $r,s,t$.

\section{Example : Higgs-boson mass in the di-photon and four-lepton channels}
\label{sec:simpleExample}

Let us consider the first ATLAS Run 2 measurement of the Higgs-boson mass, $m_\text{H}$, in the $H\to\gamma\gamma$ and $H\to 4\ell$ final states~\cite{ATLAS:2018tdk}. The measurement results in the $\gamma\gamma$ and $4\ell$ channels have similar total uncertainty, but are unbalanced in the sense that the former benefits from a large data sample but has significant systematic uncertainties from the photon energy calibration, while the latter is limited to a smaller data sample but benefits from excellent calibration systematic uncertainties:
\begin{itemize}
\item $m_{\gamma\gamma} = 124.93 \pm 0.40 (\pm 0.21 \textrm{ (stat) } \pm 0.34 \textrm{ (syst)})$~GeV;
\item $m_{4\ell} = 124.79 \pm 0.37 (\pm 0.36 \textrm{ (stat) } \pm 0.09 \textrm{ (syst)})$~GeV.
\end{itemize}

The uncertainties in the $\gamma\gamma$ and $4\ell$ measurements can be considered as entirely uncorrelated for this discussion. In the BLUE approach, the combined value and its uncertainty are then obtained considering the following log-likelihood:
\begin{equation}
    -2 \ln\mathcal{L} = \sum_{i} \left(\frac{m_i - m_\text{H}}{\sigma_{i}}\right)^2,
    \label{eq:blue}
\end{equation}
where $i=\gamma\gamma,\,4\ell$; and $\sigma_{\gamma\gamma}$~and $\sigma_{4\ell}$ are the total uncertainties in the $\gamma\gamma$ and $4\ell$ channels, respectively. The combined value $m_\text{cmb}$ and its total uncertainty $\sigma_\text{cmb}$ are derived solving
\begin{equation}
\left.\frac{\partial\ln\mathcal{L}}{\partial m_\text{H}}\right|_{m_\text{H} = m_\text{cmb}} = 0 \quad , \quad \frac{1}{\sigma_\text{cmb}^2} = \left.\frac{\partial^2\ln\mathcal{L}}{\partial m_\text{H}^2}\right|_{m_\text{H} = m_\text{cmb}}
\end{equation}
The solutions can be written in terms of linear combinations of the input values and uncertainties : 
\begin{equation}
m_\text{cmb} = \sum_i \lambda_i \, m_i
\quad , \quad
\sigma_\text{cmb}^2 = \sum_i \lambda_i^2 \, \sigma_i^2
\label{eq:wgtave}
\end{equation}
with 
\begin{equation}
\lambda_i = \frac{1/\sigma_i^2}{1/\sigma_{\gamma\gamma}^2+1/\sigma_{4\ell}^2} 
\quad , \quad
\lambda_{\gamma\gamma} + \lambda_{4\ell} = 1.
\label{eq:weightsBlue}
\end{equation}
and the weights $\lambda_i$ minimize the variance of the combined result, accounting for all sources of uncertainty in the input measurements. Since the total uncertainties have statistical and systematic components, i.e. $\sigma_i^2 = \sigma_{\text{stat},i}^2 + \sigma_{\text{syst},i}^2$, the corresponding contributions in the combined measurement are simply
\begin{equation}
\sigma_\text{stat,cmb}^2 = \sum_i \lambda_i^2 \, \sigma_{\text{stat},i}^2 
\quad , \quad
\sigma_\text{syst,cmb}^2 = \sum_i \lambda_i^2 \, \sigma_{\text{syst},i}^2.
\label{eq:uncdecompBlue}
\end{equation}

In the profile likelihood (PL) approach, or nuisance-parameter representation, the corresponding likelihood reads
\begin{equation}
    \begin{split}
        -2\ln\mathcal{L} &= \sum_{i} \left(\frac{m_i+\sum_r (\alpha_r - a_r) \Gamma_{ir} - m_\text{H}}{\sigma_{\text{stat},i}}\right)^2\\ 
        &+ \sum_r (\alpha_r - a_r)^2,
    \end{split}\label{eq:plh}
\end{equation}
where $\alpha_r$ is the nuisance parameter corresponding to the source of systematic uncertainty $r$, and $\Gamma_{ir}$ its effect on the measurement in channel $i$. Knowledge of the systematic uncertainty $r$ is obtained from an auxiliary measurement, of which the central value, sometimes called a global observable, is denoted as $a_r$. The parameters $\alpha_r$ and $a_r$ are defined in units of the systematic uncertainty $\sigma_{\text{syst},r}$, and $a_r$ is often conventionally set to 0. In this example, since the $\sigma_{\text{syst},r}$ are specific to each channel and do not generate correlations, $\Gamma_{ir} = \sigma_{\text{syst},r} \, \delta_{ir}$. The combined value $m_\text{cmb}$ and its total uncertainty are obtained from the absolute maximum and second derivative of $\mathcal{L}$ as above; in addition, the PL yields the estimated value for $\alpha_r$. One finds that $m_\text{cmb}$ and $\sigma_\text{cmb}$ exactly match their counterparts from Eq.~(\ref{eq:wgtave})~(see also the discussion in Section~\ref{sec:uncnp}).

In PL practice, the statistical uncertainty is however usually obtained by fixing all nuisance parameters to their best fit value (maximum likelihood estimator) $\hat{\alpha_r}$, maximising the likelihood only with respect to the parameter of interest. With fixed $\alpha_r$, the second derivative of Eq.~(\ref{eq:plh}) becomes equivalent to that of Eq.~(\ref{eq:blue}), only changing $\sigma_i$ for $\sigma_{\text{stat},i}$ in the denominator, giving
\begin{equation}
\sigma_\text{stat,cmb}^2 = \sum_i \lambda^{\prime 2}_{i} \, \sigma_{\text{stat},i}^2
\quad , \quad
\lambda^\prime_{i} = \frac{1/\sigma_{\text{stat},i}^2}{1/\sigma_{\text{stat},\gamma\gamma}^2+1/\sigma_{\text{stat},4\ell}^2} 
\label{eq:BLUEweights_statonly}
\end{equation}
which this time differ from Eqs.~(\ref{eq:weightsBlue}),~(\ref{eq:uncdecompBlue}): here, the coefficients $\lambda^\prime$ are calculated from the statistical uncertainties only, and optimize the combined uncertainty for this case. The statistical uncertainty is thus underestimated, comparing to Eq.~(\ref{eq:wgtave})). The systematic error, estimated from the quadratic subtraction between the total and statistical uncertainty estimate, is overestimated. 

\begin{table}[htbp]
\centering
\begin{tabular}{c|ccc}
\toprule
    Measurement   & $\sigma_\text{stat}$ & $\sigma_\text{syst}$ & $\sigma_\text{tot}$ \\
\midrule
    $\gamma\gamma$ & 0.21 & 0.34 & 0.40 \\    
    $4\ell$        & 0.36 & 0.09 & 0.37 \\    
    combined, decomposed & 0.22 & 0.16 & 0.27 \\
    combined, impacts  & 0.18 & 0.20 & 0.27 \\
    \bottomrule
\end{tabular}
\caption{Uncertainty components of $m_\text{H}$ in the $\gamma\gamma$ and $4\ell$ channels, and for the combined measurement. The combined uncertainties are given according to the BLUE result (Eq.~(\ref{eq:uncdecompBlue})) and using impacts (Eq.~(\ref{eq:BLUEweights_statonly})).\label{tab:introHiggs}}
\end{table}

For completeness, numerical values are given in Table~\ref{tab:introHiggs}. The "impact" of a systematic uncertainty on a measurement with only statistical uncertainties differs from the contribution of this systematic uncertainty to the complete measurement. In the impact procedure, the estimated measurement statistical uncertainty is actually the total uncertainty of a measurement without systematic uncertainties, i.e. of a different measurement. In other words, it does not match the standard deviation of results obtained by repeating the same measurement, including systematic uncertainties, on independent data sets of the same expected size. 

Finally, extrapolating the $\gamma\gamma$  and $4\ell$ measurements to the large data sample limit, statistical uncertainties vanish, and the asymptotic combined uncertainty should intuitively be dominated by the $4\ell$ channel and close to 0.09~GeV. A naive estimate based on impacts instead suggests an asymptotic uncertainty of 0.20 GeV.

We generalise this discussion in the following, and argue that a sensible uncertainty decomposition should match the one obtained from fits in the covariance representation, and can be also obtained simply in the context of the PL. The Higgs-boson mass example is further discussed in Section~\ref{sec:simpleExample_detailed}.

\section{Uncertainty decomposition in covariance representation}
\label{sec:unccov}

This section provides a short summary of standard results which can be found in the literature (see e.g. ~\cite{behnke_stat_textbook_chap2}). Gaussian uncertainties are assumed throughout this section. The general form of Eq.~(\ref{eq:blue}), in the presence of an arbitrary number of measurements $m_i$ and POIs $\vec{\theta}$ is:
\begin{equation}
    -2\ln\mathcal{L}_\text{cov}(\vec{\theta}) = \sum_{i,j} \left( m_i - t_i(\vec{\theta})\right) C_{ij}^{-1} \left(m_j - t_j(\vec{\theta})\right),
    \label{eq:chi2cov}
\end{equation}
where $t_i(\vec{\theta})$ are models for the $m_i$, and $C$ is the total covariance of the measurements:
\begin{equation}
    C_{ij} = V_{ij} + \sum_{r} \Gamma_{ir} \Gamma_{jr},
    \label{eq:covdef}
\end{equation}
where $V_{ij}$ represents the statistical covariance, and the second term collects all sources of systematic uncertainties. In general, $V_{ij}$ includes statistical correlations between the measurements, but is sometimes diagonal in which case $V_{ij} = \sigma_i^2 \delta_{ij}$. $\Gamma_{ir}$ represents the effect of systematic source $r$ on measurement $i$ (see Eq.~(\ref{eq:plh})), and the outer product gives the corresponding covariance.

Imposing the restriction that the models $t_i$ are linear functions of the parameters of interest, i.e. $t_i(\vec{\theta}) = t_{0,i} + \sum_p h_{ip}\theta_p$, according to the Gauss-Markov theorem~(see e.g.\ Refs.~\cite{Lyons:1988rp,Cowan:1998ji,ParticleDataGroup:2010dbb}), the POI estimators with smallest variance are found by solving $\left.\partial\ln\mathcal{L}_\text{cov}/\partial\theta_p\right|_{\vec{\theta} = \hat{\vec{\theta}}} = 0$, and the corresponding covariance is obtained from the matrix of second derivatives, $\left.\partial^2\ln\mathcal{L}_\text{cov}/\partial\theta_p\partial\theta_q\right|_{\vec{\theta} = \hat{\vec{\theta}}}$~. The solutions are
\begin{eqnarray}
    \hat{\theta}_p &=& \sum_i\lambda_{p i} (m_i - t_{0,i}), \label{eq:thetafit}\\
    \text{cov}(\hat{\theta}_p, \hat{\theta}_q) &=& \sum_{i,j} \lambda_{p i }C_{ij}\lambda_{q j},\label{eq:thetacov}
\end{eqnarray}
where the weights $\lambda_{p i}$ are given by
\begin{eqnarray}
\lambda_{p i} &=& \sum_q \left(h^{T}\cdot S\cdot h\right)_{pq}^{-1}\cdot\left(h^{T}\cdot S\right)_{q i},\label{eq:weightdef}\\
S_{ij} &=& \sum_k V_{ik}^{-1}\left(\mathbb{I} -  \Gamma \cdot Q\right)_{k j},\label{eq:Sdef}\\
Q_{r i} &=& \sum_s \left(\mathbb{I} + \Gamma^{T}V^{-1}\Gamma\right)_{rs}^{-1}(\Gamma^{T}V^{-1})_{si}.\label{eq:Qdef}
\end{eqnarray}
In particular, using Eq.~(\ref{eq:covdef}), the contribution to the uncertainties in the POIs of the statistical uncertainty in the measurements, and of each systematic source $r$, is given by 
\begin{eqnarray}
    \text{cov}^{[\text{stat}]}(\hat{\theta}_p, \hat{\theta}_q) &=& \sum_{i,j} \lambda_{p i }V_{ij}\lambda_{q j},\label{eq:thetacovstat}\\
    \text{cov}^{[r]}(\hat{\theta}_p, \hat{\theta}_q) &=& \sum_{i,j} \lambda_{p i } \left( \Gamma_{ir} \Gamma_{jr}\right)\lambda_{q j}.\label{eq:thetacovsyst}
\end{eqnarray}
We note that the BLUE averaging procedure, i.e the unbiased~\footnote{The word ``unbiased’’ employed here needs to be interpreted with care, as it actually involves several implicit assumptions about the knowledge of the input covariance matrix~(see e.g. the Chapter 7 of Ref.~\cite{Cowan:1998ji}).
Indeed, such covariances generally carry uncertainties themselves, because the size of the systematic uncertainties and their correlations are never really measured, but rather estimated.
The existence and relevance of such uncertainties on the uncertainties and on their correlations has been pointed in the context, for example, of $\alpha_{S}$ fits from jet cross section data~\cite{Malaescu:2012ts}. See also the related work in Ref.~\cite{Schmelling:1994pz}, concerning the uncertainties on uncertainties.} linear averaging of measurements of a common physical quantity, is just a special case of Eq.~(\ref{eq:chi2cov}) where the measurements are direct estimators of the POIs. In case of a single POI, $t_i=\theta$ ($t_{0,i}=0, h=1$).

A detailed discussion of template fits and of the propagation of fit uncertainties has recently been given in Ref.~\cite{Britzger:2021ocj}. While the above summary restricts to linear fits with constant uncertainties, Ref.~\cite{Britzger:2021ocj} also addresses non-linear effects and uncertainties that scale with the measured quantity, i.e. $\Gamma_{ir}\propto m_i$.

\section{Equivalence between the covariance and nuisance parameter representations}
\label{sec:uncnp}

Similarly, still assuming Gaussian uncertainties, the general form of Eq.~(\ref{eq:plh}) is:
\begin{equation}
\begin{split}
    -2\ln\mathcal{L}_\text{NP}&(\vec{\theta}, \vec{\alpha}) = \\
    &\sum_{i,j} \left(m_i - t_i(\vec{\theta}) - \sum_r \Gamma_{ir} (\alpha_r - a_r)\right)V_{ij}^{-1}\\
    &\times\left(m_j - t_j(\vec{\theta}) - \sum_s \Gamma_{js}(\alpha_s-a_s) \right)\\
    &+ \sum_r (\alpha_r - a_r)^2.
    \label{eq:chi2np}
    \end{split}
\end{equation}
The optimum of $\mathcal{L}_\text{NP}$ can be found by first minimizing Eq.~(\ref{eq:chi2np}) over $\vec{\alpha}$, for fixed $\vec{\theta}$ (i.e. \emph{profiling} the nuisance parameters $\vec{\alpha}$); substituting the result into Eq.~(\ref{eq:chi2np}) (thus obtaining the \emph{profile likelihood} $\ln\mathcal{L}_\text{NP}(\vec{\theta}, \hat{\hat{\vec{\alpha}}} (\vec{\theta})$); and minimizing over $\vec{\theta}$. The profiled nuisance parameters are given by:
\begin{equation}
    \hat{\hat{\alpha}}_r (\vec{\theta}) = \sum_i Q_{ri} \left(m_i - t_i(\Vec{\theta})\right) + a_r \label{eq:NPfit},
\end{equation}
where $Q_{ri}$ was defined in Eq.~(\ref{eq:Qdef}). The expression for the covariance is
\begin{equation}
    \text{cov}(\hat{\hat{\alpha}}_r, \hat{\hat{\alpha}}_s) (\vec{\theta}) = \left(\mathbb{I} + \Gamma^{T}V^{-1}\Gamma \right)_{rs}^{-1}.
    \label{eq:NPcov}
\end{equation}
Substituting Eq.~(\ref{eq:NPfit}) back into Eq.~(\ref{eq:chi2np}), and after some algebra, the profile likelihood can be written as
\begin{equation}
\begin{split}
    -2\ln{\cal L}_\text{NP}\left(\vec{\theta}, \hat{\hat{\vec{\alpha}}} (\vec{\theta})\right) &= \sum_{i,j} \left(m_i - t_i(\vec{\theta)}\right) S_{ij} \left(m_j - t_j(\vec{\theta)}\right),\\
\end{split}
    \label{eq:chi2npmin}
\end{equation}
where $S_{ij}$ was defined in Eq.~(\ref{eq:Sdef}). Moreover it can be verified that
\begin{eqnarray}
    \sum_k V_{ik}^{-1}\left(\mathbb{I} -  \Gamma \cdot Q\right)_{k j} &=& \left( V_{ij} + \sum_{r}\Gamma_{ir}\Gamma_{jr} \right)^{-1}, \text{  i.e}\\
    S_{ij} &=& C^{-1}_{ij},
    \label{eq:matrixS_C}
\end{eqnarray}
so that Eqs.~(\ref{eq:chi2npmin}) and~(\ref{eq:chi2cov}) are in fact identical. In other words, $\mathcal{L}_\text{cov}(\vec\theta)$, in covariance representation, can be seen as the result of maximizing $\mathcal{L}_\text{NP}(\vec\theta,\vec\alpha)$ over $\vec\alpha$, for fixed $\vec\theta$: it is the profile likelihood. Consequently, the best values for the POIs are still given by Eq.~(\ref{eq:thetafit}), their uncertainties by Eq.~(\ref{eq:thetacov}), and the error decomposition of Section~\ref{sec:unccov} applies.


The observation above is not new and has to the knowledge of the authors at least been discussed in Refs.~\cite{Demortier:1999,Fogli:2002pt,Stump:2001gu,Thorne:2002kn,Botje:2001fx,Glazov:2005rn,Barlow:2017xlo} for diagonal statistical uncertainties, and in Ref.~\cite{ListTalk,ATLAS:2013jmu} in the general case. It is also briefly mentioned in~Ref.~\cite{Britzger:2021ocj}. The equivalence between the covariance and nuisance parameter representations is reminded here to insist that profile-likelihood fits should obey the uncertainty decomposition usual from fits in the covariance representation. 

For any value of $\vec\theta$, the estimators of the nuisance parameters and their covariance are given by Eqs.~(\ref{eq:NPfit}) and~(\ref{eq:NPcov}). The estimator $\hat{\alpha}$ is given by the product of the differences between the measurements and the model, $m_i - t_i(\vec\theta)$, and a factor $Q$ determined only from the initial systematic and experimental uncertainties. This factor can be calculated from the basic inputs to the fit. Nuisance parameter pulls ($\hat{\alpha}_r$) and constraints ($\sqrt{\text{cov}(\hat{\alpha}_r, \hat{\alpha}_r)}$) can thus also be calculated $a$ $posteriori$ in the context of a POI-only fit in covariance representation, without explicitly introducing $\vec \alpha$, $\vec a$ in the expression of the likelihood, from the same inputs as those defining $C$.

This procedure can be repeated first minimizing over $\vec{\theta}$ for given $\vec{\alpha}$, substituting the result into Eq.~(\ref{eq:chi2np}), and minimising the result over the nuisance parameter $\vec{\alpha}$. This yields the NP covariance matrix elements as
\begin{equation}
    \text{cov}(\hat{\alpha}_r, \hat{\alpha}_s) = \left[\mathbb{I} + (\zeta\cdot\Gamma)^{T}V_{}^{-1}(\zeta\cdot\Gamma) \right]_{rs}^{-1},
    \label{eq:NPcov_fixedalpha}
 \end{equation}
with
 \begin{eqnarray}
    \zeta_{i j} &= & \sum_{p}h_{ip}\rho_{pj} - \delta_{ij},\\
    \rho_{pj} &=&\sum_{q}(h^T\cdot V^{-1}\cdot h)^{-1}_{pq}(h^T\cdot V^{-1})_{q j},
 \end{eqnarray}
while the covariance between the NPs and POI is given by 
\begin{equation}
     \text{cov}\left(\hat\alpha_r, \hat\theta_p\right) = -\sum_s \left[\mathbb{I} + (\zeta\cdot\Gamma)^{T}V_{}^{-1}(\zeta\cdot\Gamma) \right]_{rs}^{-1} \left(\rho\cdot\Gamma\right)_{ps}.\label{eq:cov_nps_poi}
 \end{equation}
 Equations~(\ref{eq:thetacov}),~(\ref{eq:NPcov_fixedalpha}) and~(\ref{eq:cov_nps_poi}) determine the full covariance matrix of the fitted parameters.

Importantly, Eq.~(\ref{eq:cov_nps_poi}) can be further simplified to
\begin{equation}
     \text{cov}\left(\hat\alpha_r, \hat\theta_p\right) = -\sum_i \lambda_{pi}\Gamma_{i r},\label{eq:cov_nps_poi_2}
 \end{equation}
which directly provides the systematic uncertainty decomposition. The inner product of Eq.~(\ref{eq:cov_nps_poi_2}) with itself gives the systematic covariance, Eq.~(\ref{eq:thetacovsyst}), and the statistical uncertainty can be obtained subtracting the result in quadrature from the total uncertainty in $\hat\theta_p$. In other words, the contribution of every systematic source to the total uncertainty is directly given by the covariance between the corresponding NP and the POI.

\section{Uncertainty decomposition from shifted observables}
\label{sec:UncDecomp_ShiftedObs}

While it is a common and relevant approximation, probability models are in general not based on Gaussian uncertainty distributions. Small samples are treated using the Poisson distribution, and the constraint terms associated to nuisance parameters can assume arbitrary forms. The best-fit values of the POI are however always functions of the measurements and the central values of the auxiliary measurements, i.e. $\hat\theta_p = \hat\theta_p(\vec{m},\vec{a})$. Assuming no correlations between these observables, the uncertainty in $\hat\theta_p$ then follows from linear error propagation:
\begin{eqnarray}
    \text{cov}(\hat\theta_p,\hat\theta_p) = \sum_i \left(\frac{\partial \hat\theta_p}{\partial m_i}\cdot\sigma_i\right)^2 + \sum_r \left(\frac{\partial \hat\theta_p}{\partial a_r}\cdot 1 \right)^2,
    \label{eq:thetaerrprop}
\end{eqnarray}
where $\sigma_i$ is the uncertainty in $m_i$, the uncertainty in $a_r$ is 1 by definition of $a_r$ and $\alpha_r$ (Section~\ref{sec:simpleExample}), and $\frac{\partial \hat\theta_p}{\partial m_i}$, $\frac{\partial \hat\theta_p}{\partial a_r}$ are the sensitivities of the fit result to these observables. The first sum in Eq.~(\ref{eq:thetaerrprop}) reflects the fluctuations of the measurements, i.e. the statistical uncertainty (each term of the sum represents the contribution of a given $m_i$, measurement or bin), and the second sum collects the contributions of the systematic uncertainties. 

The contribution of a given source of uncertainty can thus be assessed by varying the corresponding measurement or global observable by one standard deviation in the expression of the likelihood, and repeating the fit otherwise unchanged. The corresponding uncertainty is obtained from the difference between the values of $\hat\theta_p$ in the varied and nominal fits.

This statement can be verified explicitly for the Gaussian, linear fits discussed in the previous section. Now allowing for correlations between the measurements, varying $m_k$ within its uncertainty yields the following likelihood:
\begin{equation}
\begin{split}
    -2\ln{\cal L}_{m_k}&(\vec{\theta}, \vec{\alpha}) = \\
    &\sum_{i,j} \left(m_i + L_{ik}-t_i(\vec{\theta}) - \sum_r \Gamma_{r i} (\alpha_r-a_r) \right) V_{ij}^{-1} \\
    &\times\left( m_j + L_{jk} -t_j(\vec{\theta}) - \sum_s \Gamma_{s j} (\alpha_s - a_s)\right) \\ &+ \sum_r (\alpha_r - a_r)^2,
    \label{eq:chi2shift1}
\end{split}
\end{equation}
where $L$ results from the Cholesky decomposition $L^T L = V$, and represents the correlated effect on all measurements $m_i$ of varying $m_k$ within its uncertainty. In the case of uncorrelated measurements, $L_{ik} = \sigma_i \delta_{ik}$ and only $m_k$ is varied as in Eq.~(\ref{eq:thetaerrprop}). 
After minimization, the difference between the varied and nominal fit results is
\begin{equation}
\Delta\hat\theta^{[m_k]}_{p}\equiv
\hat\theta^{[m_k]}_{p} - \hat\theta_{p} =  \sum_{i} \lambda_{p i}L_{ik}.
\end{equation}


Similarly, the uncertainty in $a_t$ can be obtained from the following likelihood:
\begin{equation}
\begin{split}
    -2\ln{\cal L}_{a_t}&(\vec{\theta}, \vec{\alpha}) = \\
    &\sum_{i,j} \left( m_i - t_i(\vec{\theta}) - \sum_r \Gamma_{r i} (\alpha_r -a_r) \right) V_{ij}^{-1} \\
    &\times\left( m_j - t_j(\vec{\theta}) - \sum_s \Gamma_{s j} (\alpha_s -a_s) \right) \\
    &+ \sum_r (\alpha_r - a_r - \delta_{rt})^2,
    \label{eq:chi2shift2}
\end{split}
\end{equation}
resulting in
\begin{equation}
\Delta\hat\theta^{[a_t]}_{p}\equiv\hat\theta^{[a_t]}_{p} - \hat\theta_{p} =  -\sum_i \lambda_{p i}\Gamma_{it}.\label{eq:thetak}
\end{equation}
as in Eq.~(\ref{eq:cov_nps_poi_2}). The differences between the varied and nominal values of $\hat\theta_p$
match the expressions obtained above for the corresponding uncertainties. In particular,
\begin{equation}
\sum_k \Delta\hat\theta^{[m_k]}_{p}
\Delta\hat\theta^{[m_k]}_{q} = \sum_{i,j} \lambda_{p i} V_{ij} \lambda_{q j}
\label{eq:shiftstat}
\end{equation}
reproduces the total statistical covariance in Eq.~(\ref{eq:thetacovstat}), and 
\begin{equation}
\Delta\hat\theta^{[a_t]}_{p}
\Delta\hat\theta^{[a_t]}_{q} = \sum_{i,j} \lambda_{p i}\left( \Gamma_{it} \Gamma_{jt}\right) \lambda_{q j}
\label{eq:shiftsyst}
\end{equation}
is the contribution of systematic source $t$ to the systematic covariance in Eq.~(\ref{eq:thetacovsyst}).

As in Section~\ref{sec:uncnp}, the total uncertainty in the NPs can be obtained minimizing the likelihood with respect to $\vec\theta$ for fixed $\vec\alpha$, replacing $\vec\theta$ by its expression, and minimizing the result with respect to $\vec\alpha$. The contribution of the measurements to the uncertainty in $\vec\alpha$ is
\begin{equation}
    \Delta\hat\alpha^{[m_k]}_r=\hat\alpha^{[m_k]}_r - \hat\alpha_r =\sum_i \tilde Q_{ri} L_{ik},\label{eq:nps_stat_fluct}
\end{equation}
where
\begin{equation}
    \tilde Q_{ri} = -\sum_s \left[\mathbb{I} + (\zeta\cdot\Gamma)^{T}V^{-1}(\zeta\cdot\Gamma) \right]_{rs}^{-1} \left[(\zeta\cdot \Gamma)^T\cdot V^{-1}\right]_{si};
\end{equation}
and the systematic contributions are given by
\begin{equation}
    \Delta\hat\alpha^{[a_t]}_r=    \hat\alpha^{[a_t]}_r - \hat\alpha_r = \left[\mathbb{I} + (\zeta\cdot\Gamma)^{T}V^{-1}(\zeta\cdot\Gamma) \right]_{rt}^{-1}.\label{eq:nps_sys_fluct}
\end{equation}
Summing Eqs.~(\ref{eq:nps_stat_fluct}) and~(\ref{eq:nps_sys_fluct}) in quadrature recovers the total NP covariance matrix in Eq.~(\ref{eq:NPcov_fixedalpha}), as expected.

Finally, the covariance between the NPs and POIs can be obtained analytically by summing the products of the corresponding offsets, obtained from statistic and systematic variations, that is, 
\begin{equation}
\begin{split}
    \sum_k \Delta\alpha^{[m_k]}_r &\Delta\theta^{[m_k]}_p + \sum_t \Delta\alpha^{[a_t]}_r \Delta\theta^{[a_t]}_p\\
    &= - \sum_s \left[\mathbb{I} + (\zeta\cdot\Gamma)^{T}V^{-1}(\zeta\cdot\Gamma) \right]_{rs}\left(\rho\cdot\Gamma\right)_{ps},
\end{split}\label{eq:np_poi_cov}
\end{equation} 
which again matches the expression for $\text{cov}(\hat\alpha_r,\hat\theta_p)$ in Eq.~(\ref{eq:cov_nps_poi}).

The identities~\ref{eq:shiftstat}, \ref{eq:shiftsyst}, \ref{eq:nps_sys_fluct}, \ref{eq:np_poi_cov} can be obtained analytically only in the context of Gaussian uncertainties, but the uncertainty decomposition through fits with shifted observables results from the Taylor expansion of Eq.~(\ref{eq:thetaerrprop}) and is therefore general. While analytical fits can use the representation that is the most practical for their particular purpose, numerical profile likelihood fits, assuming Poisson statistics and/or non-Gaussian nuisance parameter distributions, can still rely on Eq.~(\ref{eq:thetaerrprop}) to obtain a consistent uncertainty decomposition where each component directly reflects the impact of fluctuations in the corresponding source to the total variance of the measurement. In this way, uncertainty components preserve a universal meaning, regardless of the statistical method used for a given measurement.

In practice, the uncertainty can be propagated using one-standard-deviation shifts in $m$ and $a$ as above, or using the Monte Carlo error propagation method, where $m$ or $a$ are randomized within their respective probability density functions, and the corresponding uncertainty in the measurement is determined from the variance of the fit results.~\footnote{In order to perform the uncertainty propagation in a linear regime, one can also apply shifts of less than one-standard-deviation, followed by a rescaling of the resulting propagated uncertainty. For effectively probing possible non-linear effects impacting the tails of the uncertainty distributions, one can perform a scan of the shifts by e.g. $1$, $2$, ... $5$ standard deviations.}
The latter method makes the correspondence between uncertainty contributions and the effect of fluctuations of the corresponding sources (cf. Section~\ref{sec:intro}) explicit. It is also more general, and gives more precise results in case of significant asymmetries or tails in the uncertainty distributions. It can also be more efficient, when simultaneously estimating the variance contributed by a large group of sources of uncertainty. Similarly, the present method can be generalized to unbinned measurements using data resampling techniques for the extraction of statistical uncertainty components~\cite{Efron1992}.

\section{Examples}
\label{sec:examples}
\subsection{Combination of two measurements} 
\label{sec:simpleExample_detailed}
Let us consider again the concrete case of the Higgs boson mass $m_\text{H}$ described in Section~\ref{sec:simpleExample}, which will serve as a simple example with only one parameter of interest ($m_\text{H}$) and two measurements. We will further assume that both the statistical and systematic uncertainties are uncorrelated between the two channels, which is not unreasonable given that they correspond to different events and that the dominating sources of systematic uncertainty are indeed uncorrelated. We will take numerical values from the actual ATLAS~\cite{ATLAS:2018tdk} and CMS~\cite{CMS:2020xrn} Run~1 and Run~2 measurements, as well as from an imaginary case exaggerating the numeric features of the ATLAS Run~2 measurement.

For each case, the decomposition of uncertainties between statistical and systematic components will be compared between the two approaches -- uncertainty decomposition and impacts. In addition, this is done as a function of a luminosity factor $k$, which is used to scale the statistical uncertainty of the inputs by $1/\sqrt{k}$ (while systematic uncertainties are kept unchanged). The published results in the example under consideration are for $k=1$. Though not shown on the plots, we have also checked numerically that the uncertainty decomposition (as usually done in covariance representation methods or BLUE) can be reproduced from a profile likelihood fit with shifted observables (Section \ref{sec:uncnp}), while the impacts (as usually done in profile likelihood fits) can also be recovered from the BLUE approach, simply by using the statistical uncertainties alone to compute the combination weights $\lambda^\prime_i$ as in Eq.~(\ref{eq:BLUEweights_statonly}) (i.e. repeating the combination without systematic uncertainties). In addition, both approaches have been checked to yield to the same total uncertainty in all cases.

\begin{figure*}[htbp]
    \centering
    \subfloat[]{\includegraphics[width=0.4\textwidth]{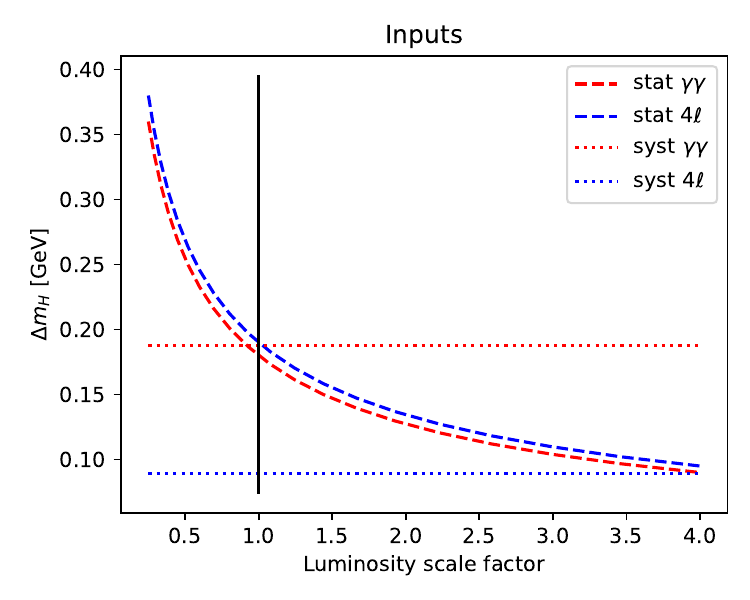}}
    \subfloat[]{\includegraphics[width=0.4\textwidth]{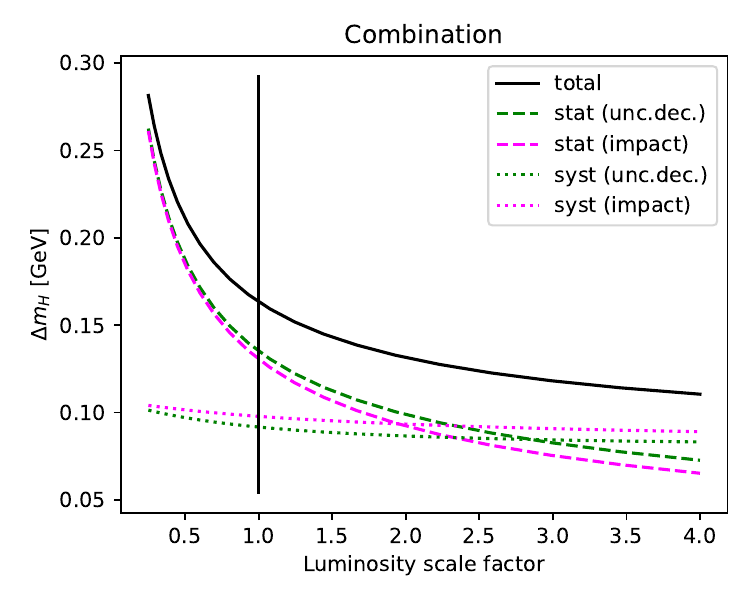}}
    \caption{Uncertainty decomposition as a function of a luminosity scaling factor, using CMS Run~2 results~\cite{CMS:2020xrn}. Left: size of the statistical (stat) and systematic (syst) uncertainties for $\gamma\gamma$ and $4\ell$. Right: decomposition of uncertainties on the combination using either the uncertainty decomposition or impacts approach.}
    \label{fig:mhcomb_cmsrun2}
\end{figure*}

\paragraph{CMS results}
We first study the combination of CMS Run~2 results~\cite{CMS:2020xrn}: $\text{stat}_{\gamma\gamma} = 0.18$\,GeV, $\text{syst}_{\gamma\gamma} = 0.19$\,GeV; $\text{stat}_{4\ell} = 0.19$\,GeV, $\text{syst}_{4\ell} = 0.09$\,GeV. The results of our toy combination are shown in Fig.~\ref{fig:mhcomb_cmsrun2}. This figure, as well as the following ones, comprises two panels: the inputs to the combination on the left, and statistical and systematic uncertainties as obtained in either the uncertainty decomposition or impact approaches on the right. The actual published numbers~\cite{CMS:2020xrn} correspond to $k=1$ (black vertical line).

With this first simple case, where the two measurements have relatively comparable uncertainties, little difference is found between the two approaches, though the uncertainty decomposition gives a larger statistical uncertainty than the impact one, as expected. The difference becomes larger for higher values of the luminosity factor. 

\begin{figure*}[htbp]
    \centering
    \subfloat[]{\includegraphics[width=0.4\textwidth]{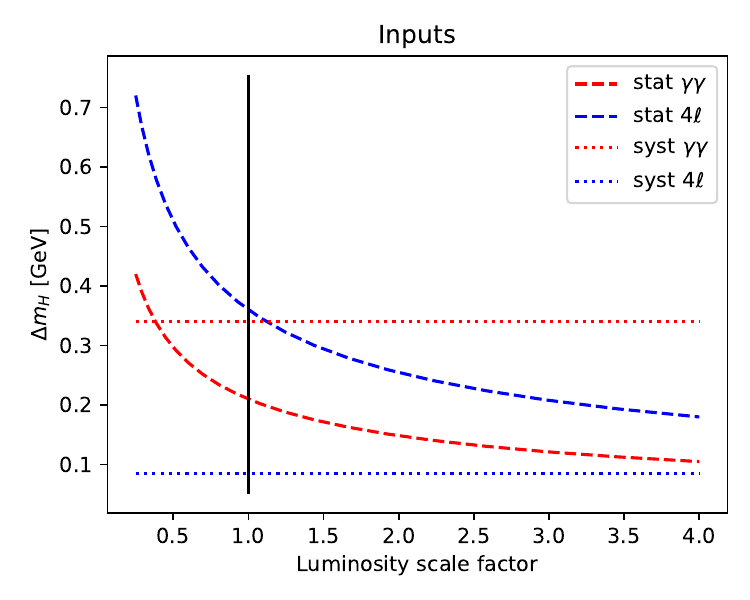}}
    \subfloat[]{\includegraphics[width=0.4\textwidth]{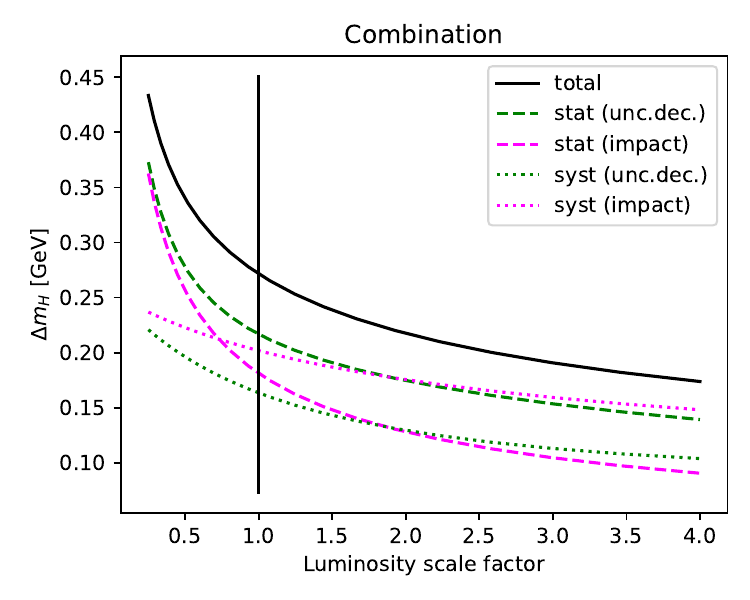}}
    \caption{Uncertainty decomposition as a function of a luminosity scaling factor, using ATLAS Run~2 results~\cite{ATLAS:2018tdk}. Left: size of the statistical (stat) and systematic (syst) uncertainties for $\gamma\gamma$ and $4\ell$. Right: decomposition of uncertainties on the combination using either the uncertainty decomposition or impacts approach.}
    \label{fig:mhcomb_atlasrun2}
\end{figure*}

\paragraph{ATLAS results}
We are now considering the ATLAS Run~2 results~\cite{ATLAS:2018tdk}: 
 $\text{stat}_{\gamma\gamma} = 0.21$\,GeV, $\text{syst}_{\gamma\gamma} = 0.34$\,GeV; $\text{stat}_{4\ell} = 0.36$\,GeV, $\text{syst}_{4\ell} = 0.09$\,GeV. As shown in Fig.~\ref{fig:mhcomb_atlasrun2}, differences between the two uncertainty decompositions are now more evident, already for the nominal uncertainty but even more when extrapolating to larger luminosities (smaller statistical uncertainties). Again, the uncertainty decomposition gives a large statistical uncertainty than the impact one.

\begin{figure*}
    \centering
    \subfloat[]{\includegraphics[width=0.4\textwidth]{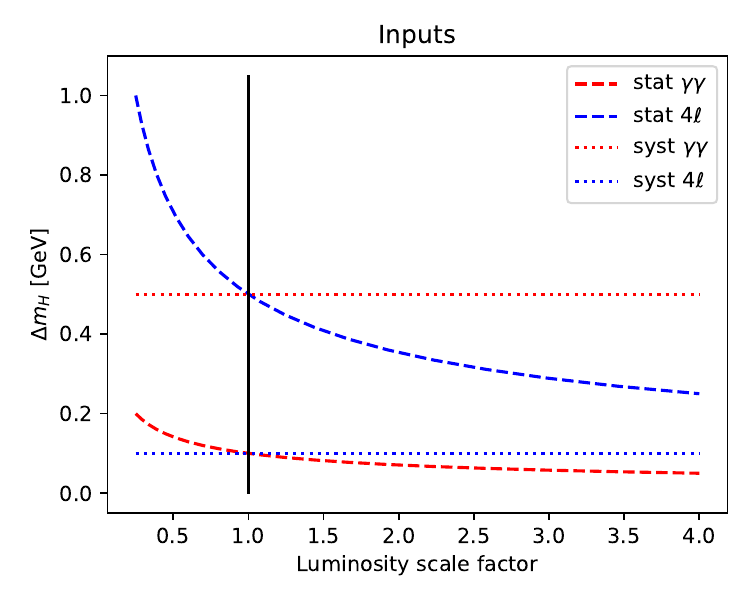}}
    \subfloat[]{\includegraphics[width=0.4\textwidth]{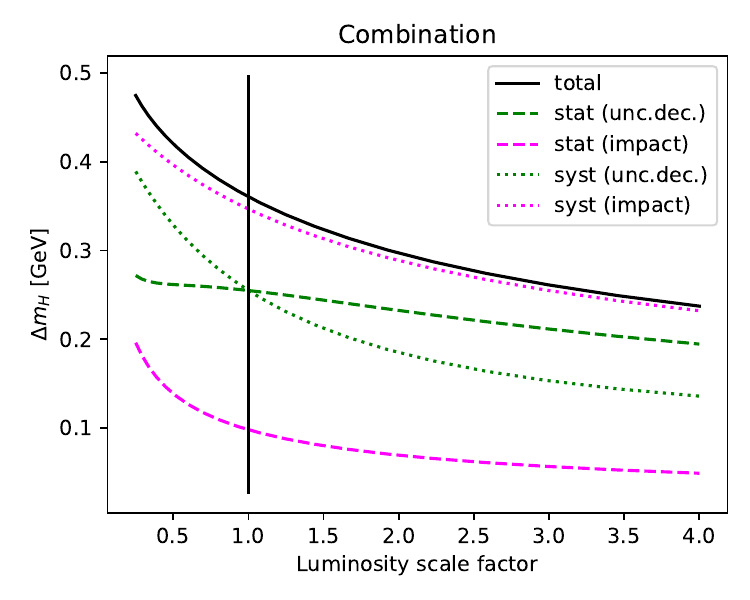}}
    \caption{Uncertainty decomposition as a function of a luminosity scaling factor, using $\text{stat}_{\gamma\gamma} = 0.1$\,GeV, $\text{syst}_{\gamma\gamma} = 0.5$\,GeV; $\text{stat}_{4\ell} = 0.5$\,GeV, $\text{syst}_{4\ell} = 0.1$\,GeV. Left: size of the statistical (stat) and systematic (syst) uncertainties for $\gamma\gamma$ and $4\ell$. Right: decomposition of uncertainties on the combination using the uncertainty decomposition or impact approach.}
    \label{fig:mhcomb_extreme}
\end{figure*}

\paragraph{Imaginary extreme case}
Finally, we consider an extreme case, such that $\text{stat}_{\gamma\gamma} = 0.1$\,GeV, $\text{syst}_{\gamma\gamma} = 0.5$\,GeV; $\text{stat}_{4\ell} = 0.5$\,GeV, $\text{syst}_{4\ell} = 0.1$\,GeV, exaggerating the features of the ATLAS combination (i.e., combining a statistically-dominated measurement with a systematically-limited one). Dramatic differences between the two approaches for uncertainty decomposition are observed in Fig.~\ref{fig:mhcomb_extreme}: for the nominal luminosity, while uncertainty decomposition reports equal statistical and systematic uncertainties, the impacts are dominated by the systematic uncertainty.

\subsection{$W$-boson mass fits}

The uncertainty decomposition discussed above is further illustrated with a toy measurement of the $W$-boson mass using pseudo-data, where the results obtained from the profile likelihood fit and from the analytical calculation are compared. Since the measurement of W mass is a typical shape analysis, in which the fit to the distributions is parameterized by both POI and NPs, the conclusions drawn from this example can in principle be generalized to all kinds of shape analyses. While the effect of varying the W mass is parameterized by the POI, three representative systematic sources of a W mass measurement at hadron colliders~\cite{D0:2012kms,ATLAS:2017rzl,LHCb:2021bjt,CDF:2022hxs} are parameterized by NPs in the probability model: the lepton momentum scale uncertainty, the hadronic recoil (HR) resolution uncertainty and the $p_\text{T}^W$ modelling uncertainty. The W mass
is extracted from the $p_\text{T}^\ell$ or $m_\text{T}$ spectra, since measurements based on these two distributions have very different sensitivities to certain types of systematic uncertainties.

\subsubsection*{Simulation}
The signal process under consideration is the charged-current Drell-Yan process~\cite{PhysRevLett.25.316} $p p\to W^{-}\to\mu^{-}\nu$ at a center-of-mass energy of $\sqrt{s}=$13~TeV, generated using Madgraph, with initial and final state corrections obtained using Pythia8~\cite{Madgraph:2014,Pythia:2008}. Detailed information of the event generation is listed in Table \ref{tab:mc_samples}.

\begin{table}[htpb]
    \centering
    \begin{tabular}{c|c}\hline
       \multicolumn{2}{c}{Event Generator} \\\hline
       \multicolumn{2}{c}{$pp\rightarrow W^- \rightarrow \mu^- \nu_\mu$ at $\sqrt{s}$=13 [TeV]}\\\hline
        Number of events & 10,000,000\\
        Matrix elements & Madgraph at LO\\
        Input $m_W$ & 80.419 [GeV]\\
        Input $\Gamma_W$ & 2.0476 [GeV] \\
        Parton shower \& QED FSR & Pythia8 \\
    \hline
    \end{tabular}
    \caption{Madgraph+Pythia8~\cite{Madgraph:2014,Pythia:2008} event generation for MC samples. Events with an off-shell boson are excluded in the event generation at parton level, leading to a total cross-section of 6543~pb.}
    \label{tab:mc_samples}
\end{table}

Kinematic distributions for different values of the W mass 
 are obtained in simulation via Breit-Wigner reweighting~\cite{BARDIN1988539}. The systematic variations of $p_\text{T}^W$ are implemented using a linear reweighting as a function of $p_\text{T}^W$ before event selection, then taking only the shape effect on the underlying $p_\text{T}^W$ spectrum.

At reconstruction level, the $p_\text{T}$ of the bare muon is smeared by 2\% following a Gaussian distribution. A source of systematic uncertainty in the calibration of the muon momentum scale is considered. The hadronic recoil $\vec{u}_\text{T}$ is taken to be the opposite of $\vec{p}_\text{T}^W$ and smeared by a constant 6~GeV in both directions of the transverse plane. The second source of experimental systematic is taken to be the uncertainty in the calibration of the hadronic recoil resolution.
The information about the W mass templates and the systematic variations is summarized in Table \ref{tab:mc_templates}.

\begin{table}[htpb]
    \centering
    \begin{tabular}{c|c}\hline
       \multicolumn{2}{c}{Templates and systematic variations} \\\hline
        W mass templates & $\pm$ 50 MeV by Breit-Wigner reweighting\\
        $\mu$ calib. & Muon momentum scale $\pm 0.5$\textperthousand\\\
        HR calib. & Recoil resolutions $\pm 5$\textperthousand\\\
        $p_\text{T}^W$ model  & \multirow{2}{*}{$w(p_\text{T}^W) = 0.96 + 8\cdot 10^{-4} \cdot p_\text{T}^W$ [GeV]}\\ 
        (Reweighting) & \\\hline
    \end{tabular}
    \caption{W mass templates and systematic variations for the Madgraph+Pythia8 samples.}
    \label{tab:mc_templates}
\end{table}

\begin{figure*}
\centering
\subfloat[]{\includegraphics[width = 7cm, height = 7cm]{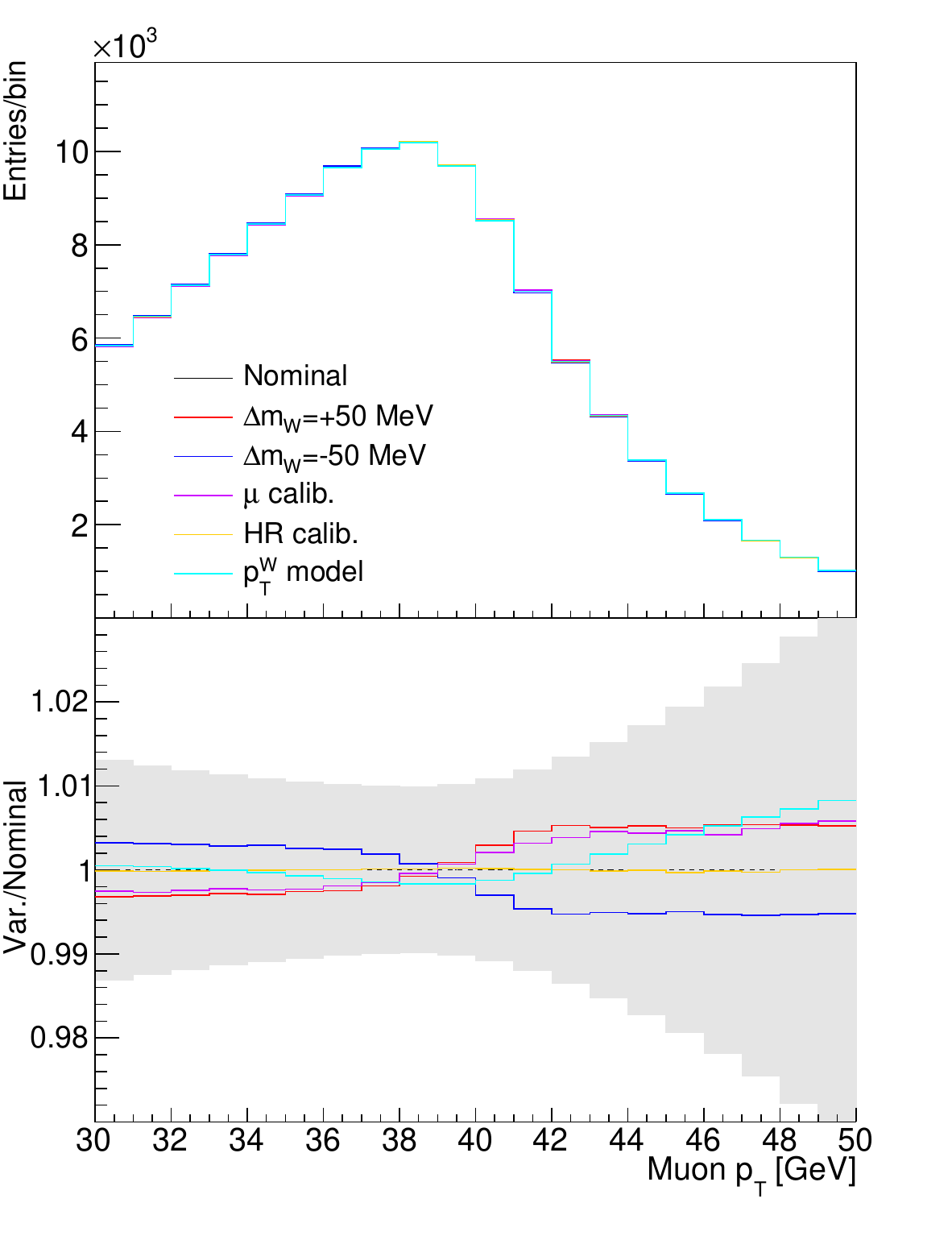}}
\subfloat[]{\includegraphics[width = 7cm, height = 7cm]{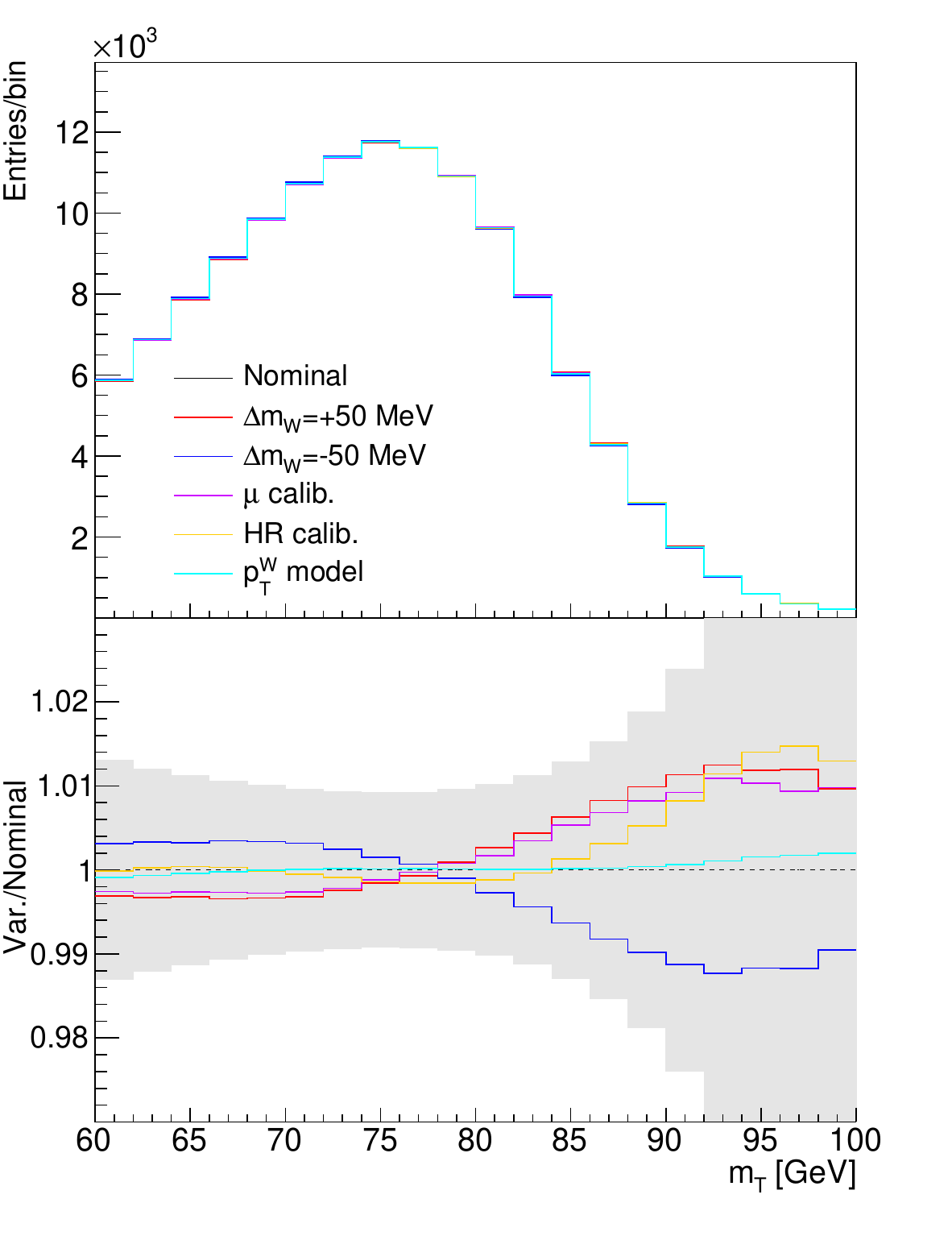}}
\caption{Reconstructed muon $p_\text{T}$ and $m_\text{T}$ distributions of the Madgraph + Pythia8 samples. (top): Kinematic spectra. (bottom): The variation to nominal ratio with statistical uncertainty indicated by the error band.}
\label{fig:mc_samples}
\end{figure*}

The detector smearing, as well as the event selections listed in Table \ref{tab:mc_samples_reco}, are chosen to be similar to those of a realistic W mass measurement. The reconstructed muon $p_\text{T}$ and $m_\text{T}$ spectra in the fit range after the event selection are shown in Figure~\ref{fig:mc_samples}, along with the relevant templates and systematic variations.

\begin{table}[htpb]
    \centering
    \begin{tabular}{c|c}\hline
        \multicolumn{2}{c}{ Detector smearing} \\\hline
        Lepton $p_\text{T}$ resolution & $2\%$ \\
        Nominal recoil resolutions & 6 [GeV] \\\hline
        \multicolumn{2}{c}{ Event selection} \\\hline
        $\eta_\ell$ selection & [-2.5, 2.5]\\
        $p_\text{T}^\ell$ selection & >25 [GeV]\\
        $E_\text{T}^{\text{miss}}$ selection & >25 [GeV] \\
        $m_\text{T}$ selection & >50 [GeV]\\
        $u_\text{T}$ selection & <25 [GeV] \\\hline
    \end{tabular}
    \caption{ Detector smearing and event selection for Madgraph+Pythia8 samples. The cut-flow efficiency of the event selection is about 29\%.}
    \label{tab:mc_samples_reco}
\end{table}

\subsubsection*{Uncertainty decomposition}
The profile likelihood fit is performed using HistFactory~\cite{HistFactory:2012} and RooFit~\cite{RooFit:2003}. Its output includes the fitted central values and uncertainties for all the free parameters. The uncertainty components of the profile likelihood fit results are obtained by repeating the fit to bootstrap samples obtained by resampling the pseudo data used to compute the results, or those of the central values of the auxiliary measurements, then computing the spread of offsets in the POI, the analytical solution of the fit can be calculated following the procedures in Section~\ref{sec:UncDecomp_ShiftedObs}. For this exercise, the pseudo data is chosen to be the nominal simulation, but with the statistical power of the data. The effect of changing the luminosity scale factor is emulated by repeating the fit with an overall factor multiplied to all the reconstructed distributions. The setups of the fits for the validation are summarized in Table~\ref{tab:setup_PLHfit}.

\begin{table*}
    \centering
    \begin{tabular}{c|c|c}\hline
        Probability model & Full model & Stat. only (for impact approach) \\\hline
         & $\mu$ calib. & \\
         NPs & HR calib. & ---\\
         & $p_\text{T}^W$ model & \\\hline
        Luminosity scale factor  & \multicolumn{2}{c}{0.1, 0.4, 0.7, 1.0, 2.0, 5.0, 10.0} \\\hline
        Fit range & \multicolumn{2}{c}{30<$p_\text{T}^\ell$<50 [GeV]}\\
         & \multicolumn{2}{c}{60<$m_\text{T}$<100 [GeV]} \\\hline
    \end{tabular}
    \caption{ Configuration of the $m_W$ fits. The luminosity scale factor of 1.0 corresponds to 76.42~[$\text{pb}^{-1}$].}
    \label{tab:setup_PLHfit}
\end{table*}

\begin{figure*}
\centering
\subfloat[Total uncertainty.]
{\includegraphics[width=0.4\textwidth]{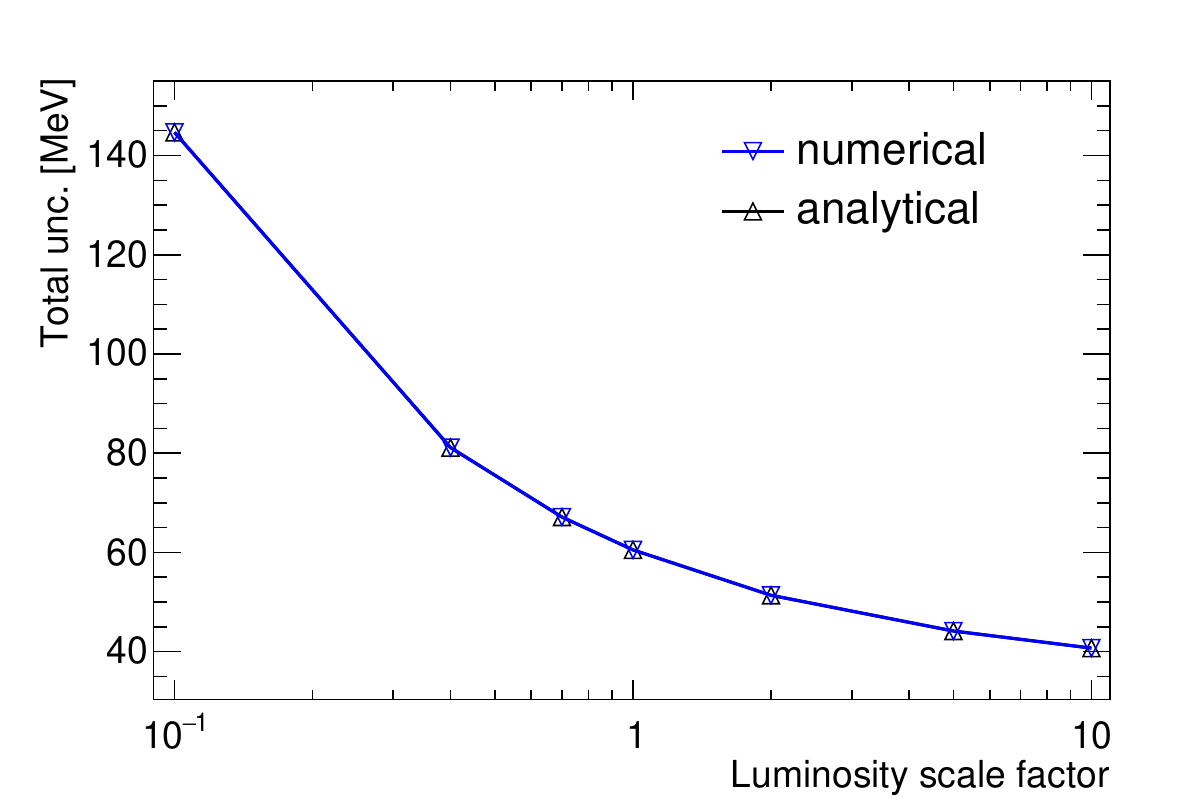}}
\subfloat[Statistical uncertainty.]{\includegraphics[width=0.4\textwidth]{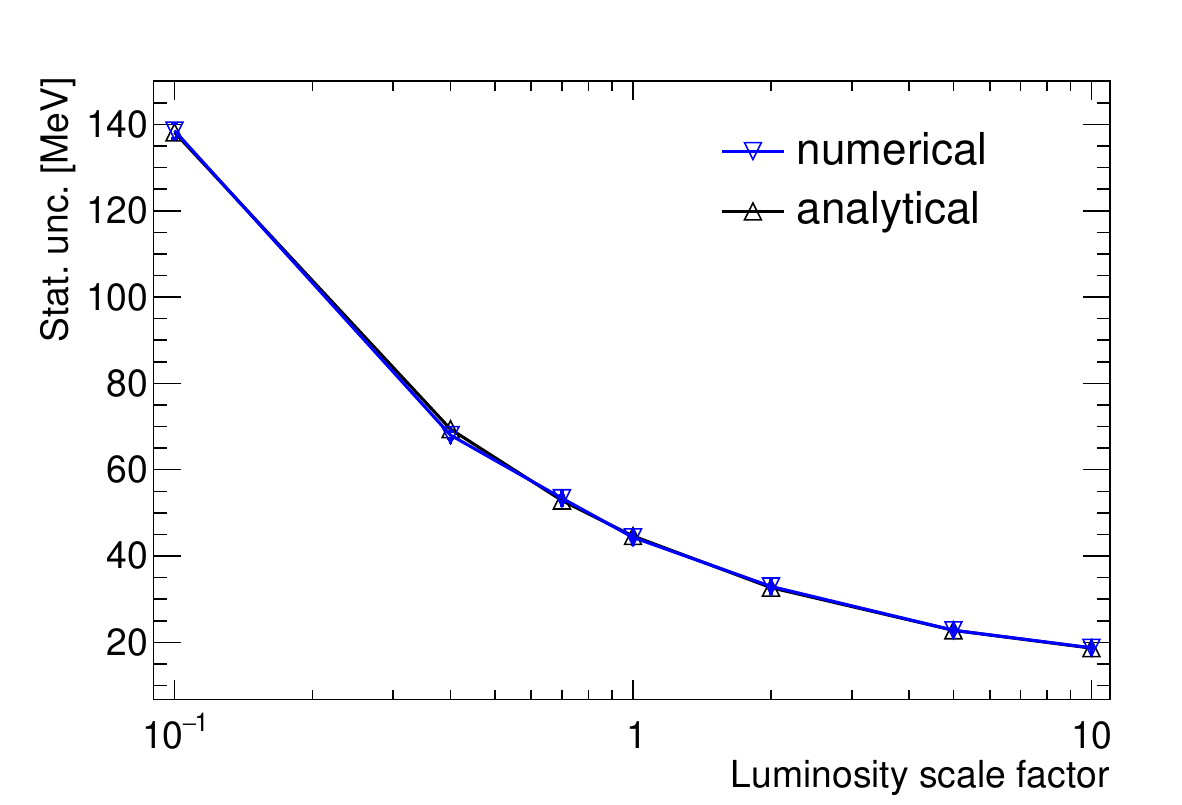}}\\
\subfloat[Total systematic uncertainty.]{\includegraphics[width=0.4\textwidth]{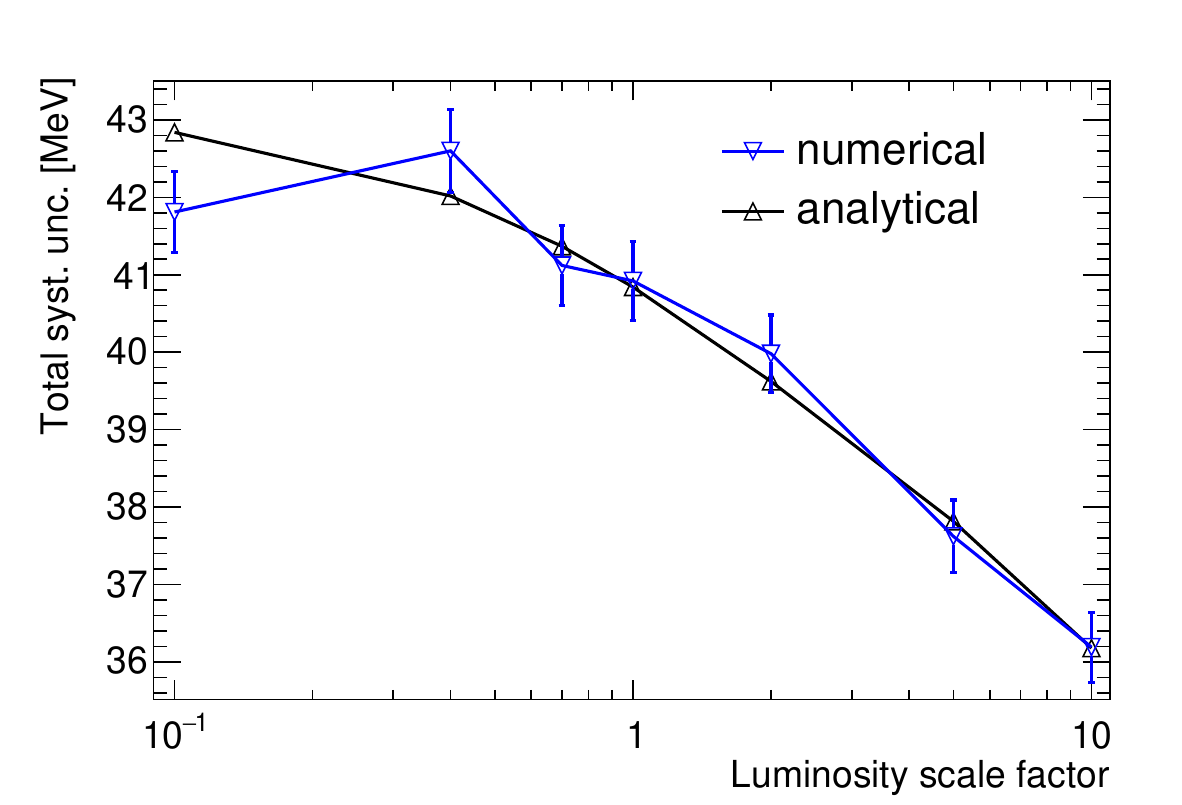}}
\subfloat[Breakdown of systematic uncertainties.]{\includegraphics[width=0.4\textwidth]{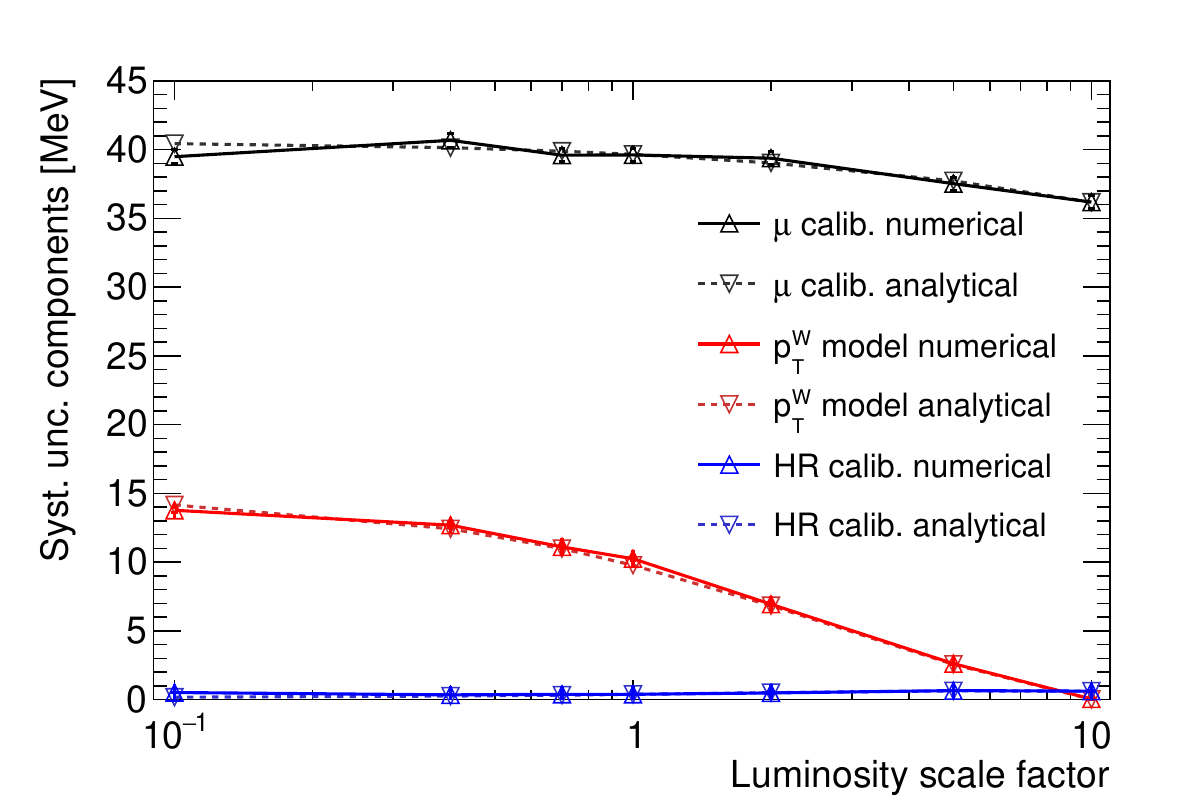}}
\caption{Uncertainty decomposition for the muon $p_\text{T}$ fit compared between the numerical and the analytical PL fit. 
 The total systematic uncertainty of the profile likelihood fit is the quadratic sum of the three components.}
\label{fig:UncDecomp_ptl}
\end{figure*}

\begin{figure*}[t]
\centering
\subfloat[Total uncertainty.]{\includegraphics[width=0.4\textwidth]{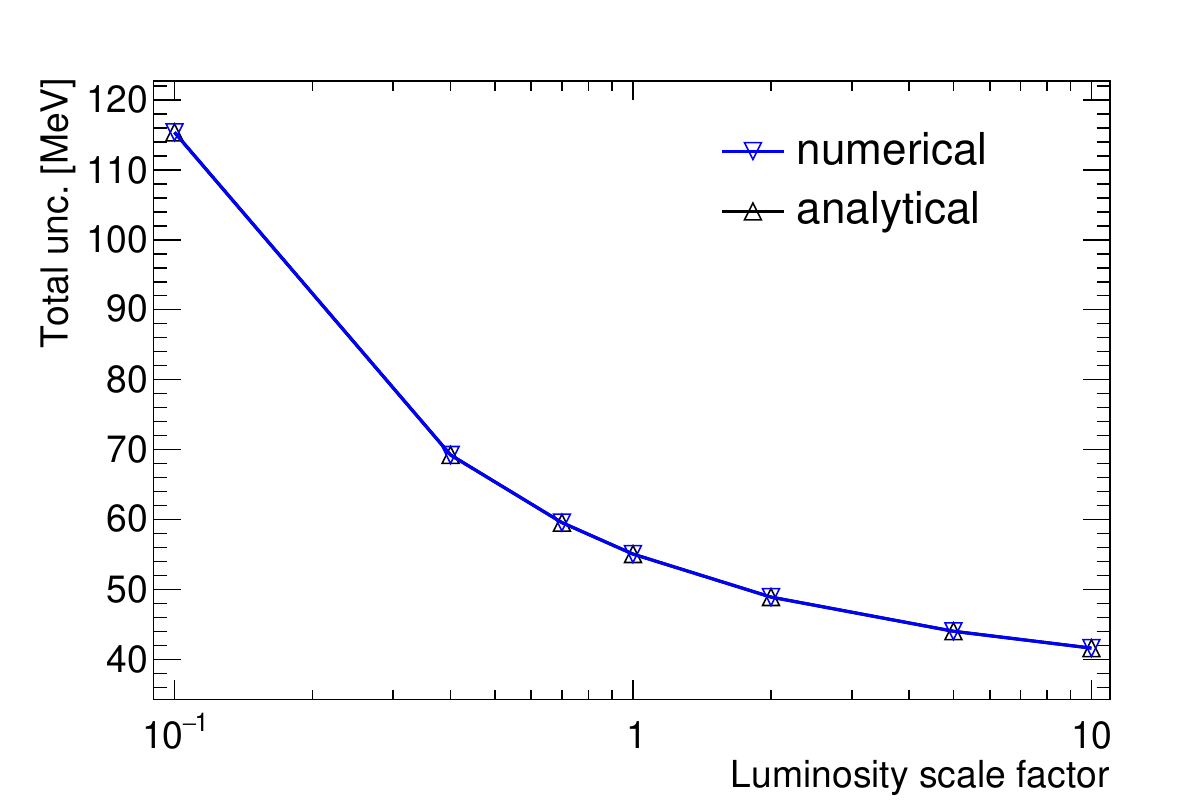}}
\subfloat[Statistical uncertainty.]{\includegraphics[width=0.4\textwidth]{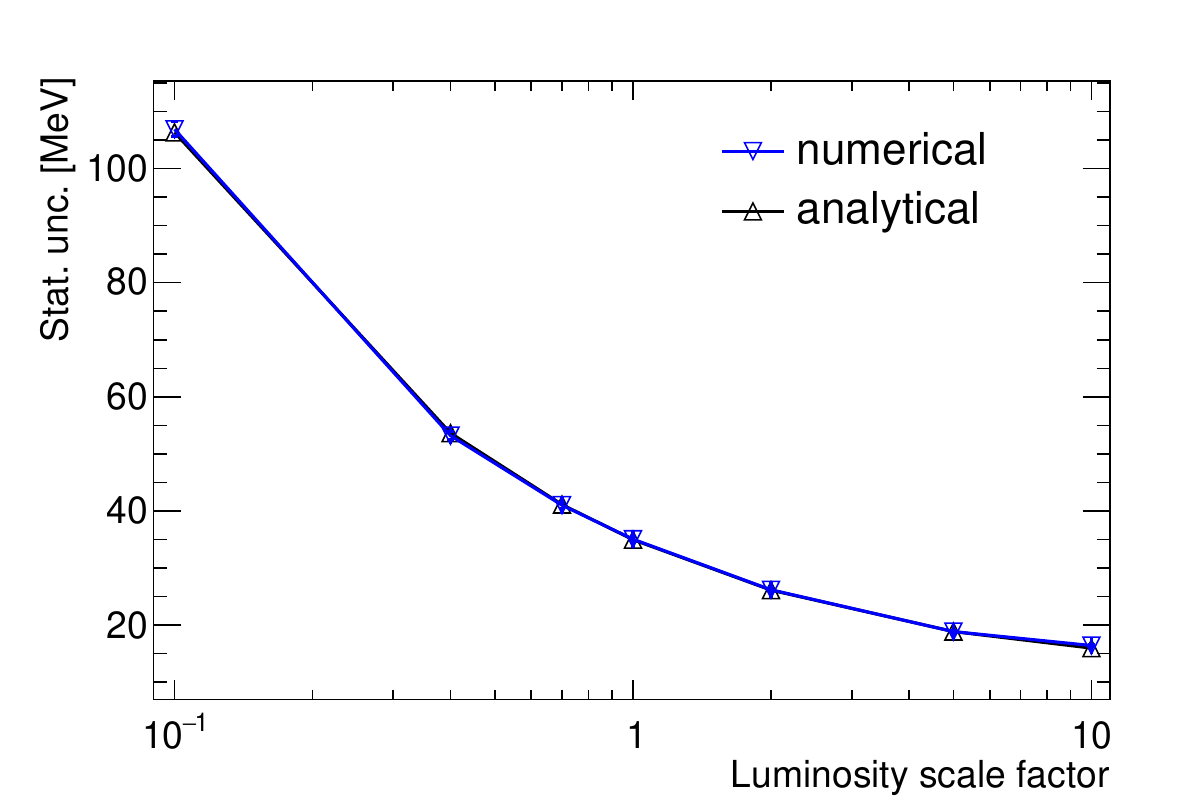}}\\
\subfloat[Total systematic uncertainty.]{\includegraphics[width=0.4\textwidth]{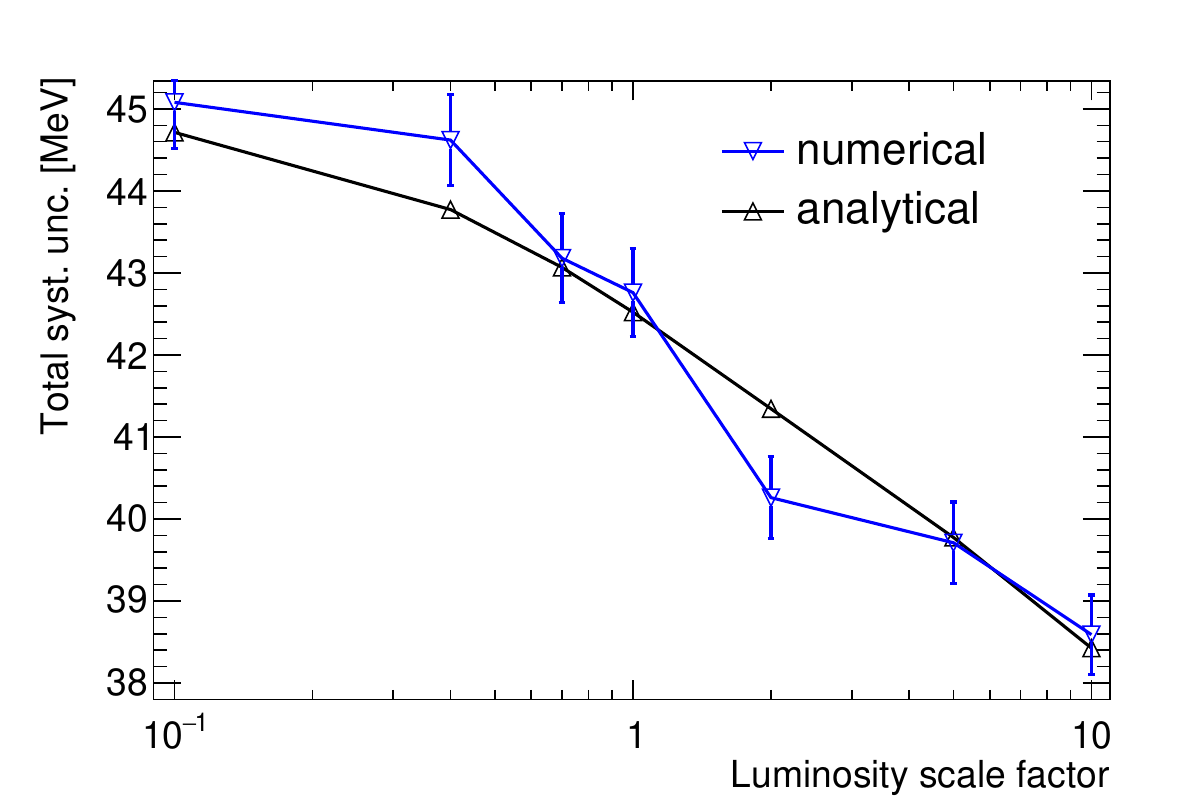}}
\subfloat[Breakdown of systematic uncertainties.]{\includegraphics[width=0.4\textwidth]{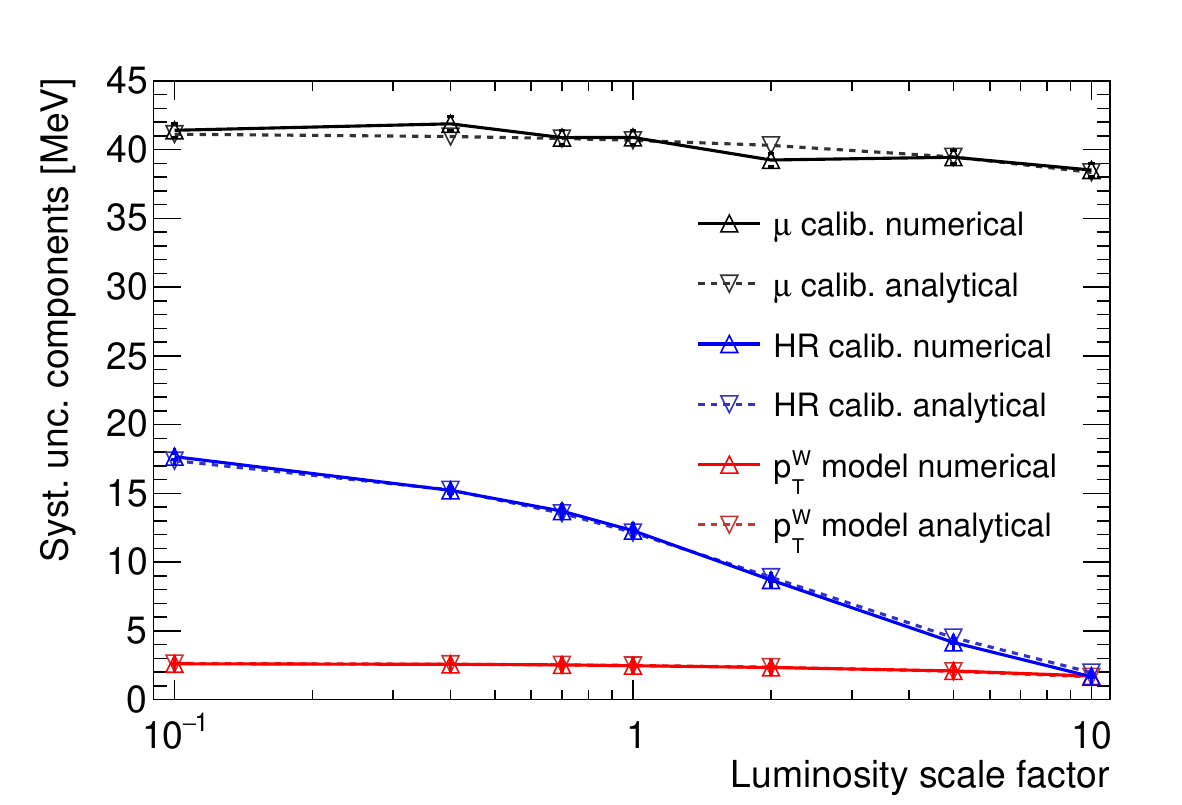}}
\caption{Uncertainty decomposition for the $m_\text{T}$ fit compared between the numerical and the analytical PL fit.
 The total systematic uncertainty of the profile likelihood fit is the quadratic sum of the three components.}
\label{fig:UncDecomp_mT}
\end{figure*}

Figures~\ref{fig:UncDecomp_ptl} and \ref{fig:UncDecomp_mT} present the uncertainty decomposition as a function a luminosity scale factor used to scale the statistical precision of the simulated sample. The error bars for the uncertainty decomposition for the profile likelihood fit reflect the limited number of toys. In general, the uncertainty components derived from the numerical profile likelihood fit and the analytical solution match each other within the error bars. The discrepancy at certain points can be assigned to the numerical stability of the PL fit, which shows up when the uncertainty components becomes too small (typically < 2~MeV). The uncertainty decomposition is summarized in Table~\ref{tab:ptlmt_UncDecomp_comparison}, where the total uncertainty is broken down into data statistic and total systematic uncertainties using the shifted observable method, and compared with the results using the conventional impact approach for PL fit. With 10 times higher luminosity, the statistical uncertainty of the impact approach decreases by exactly a factor of $\sqrt{10}$, while that of the shifted observable approach introduced in this study decreases slower.

\begin{table*}
    \centering
    \begin{tabular}{cc|ccc|ccc}\hline
        & & \multicolumn{3}{c|}{$p_\text{T}^\ell$ fit unc. [MeV]} & \multicolumn{3}{c}{$m_\text{T}$ fit unc. [MeV]} \\\hline
        Lumi & Method & $\sigma_{stat}$ & $\sigma_{syst}$ & $\sigma_{tot}$ & $\sigma_{stat}$ & $\sigma_{syst}$ & $\sigma_{tot}$ \\\hline
        $\times$  1 & shifted obs. & 44.0 $\pm$ 0.6 & 41.0 $\pm$ 1.0 & 60.5  & 34.4 $\pm$ 0.4 & 39.1 $\pm$ 1.0 & 55.1 \\
        $\times$  1 & impact & 43.7 & 41.9 & 60.5  & 33.6 & 43.6 & 55.1 \\
        $\times$ 10 & shifted obs. & 18.8 $\pm$ 0.2 & 36.2 $\pm$ 0.5 & 40.7 & 15.7 $\pm$ 0.2 & 38.6 $\pm$ 0.5 & 41.6 \\
        $\times$ 10 & impact & 13.8 & 38.3 & 40.7 & 10.6 & 40.2 & 41.6 \\\hline
    \end{tabular}
    \caption{Uncertainty decomposition for the muon $p_\text{T}^\ell$ and $m_\text{T}$ fits, for two different values of the luminosity scale factor, using the shifted observable method and the impact method for PL fit. The errors arise from the limited number of bootstrap toys. The baseline luminosity is 76.42~[$\text{pb}^{-1}$].}
    \label{tab:ptlmt_UncDecomp_comparison}
\end{table*}

\begin{figure*}[t]
\centering
\subfloat[Muon $p_\text{T}$ fit.]{\includegraphics[width=0.4\textwidth]{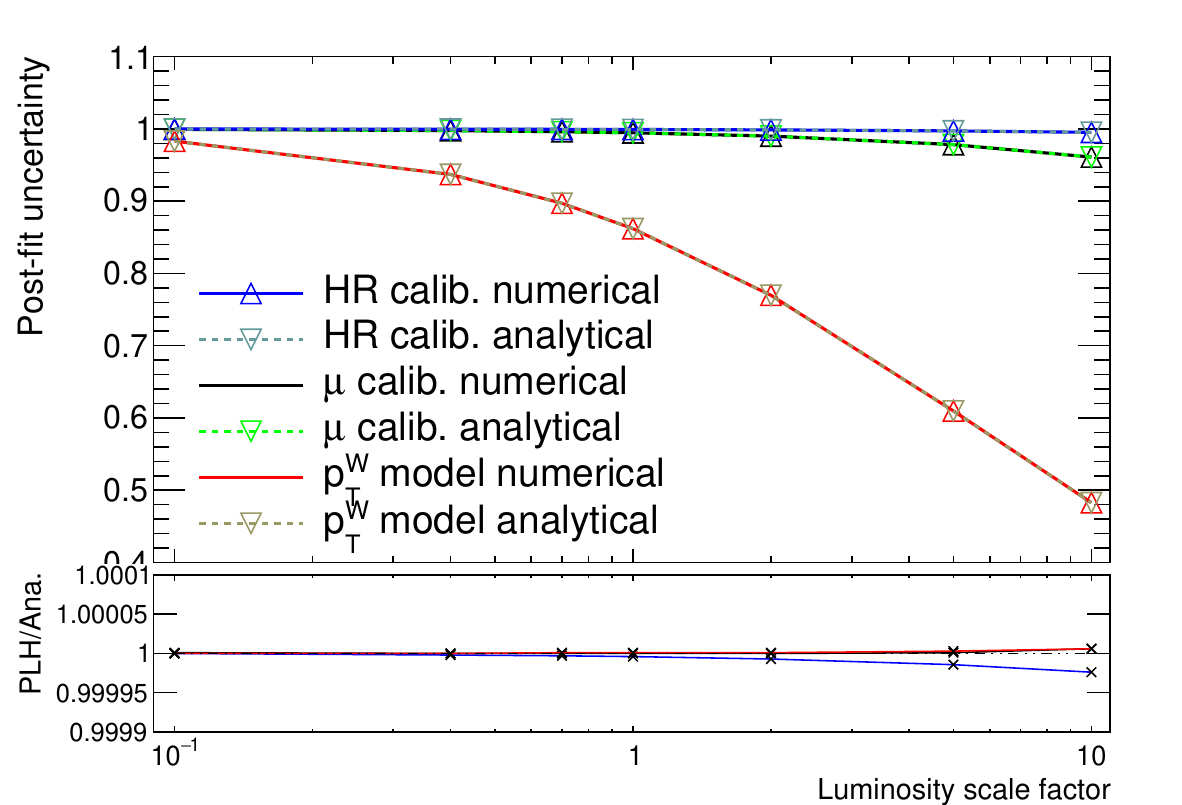}}
\subfloat[$m_\text{T}$ fit.]{\includegraphics[width=0.4\textwidth]{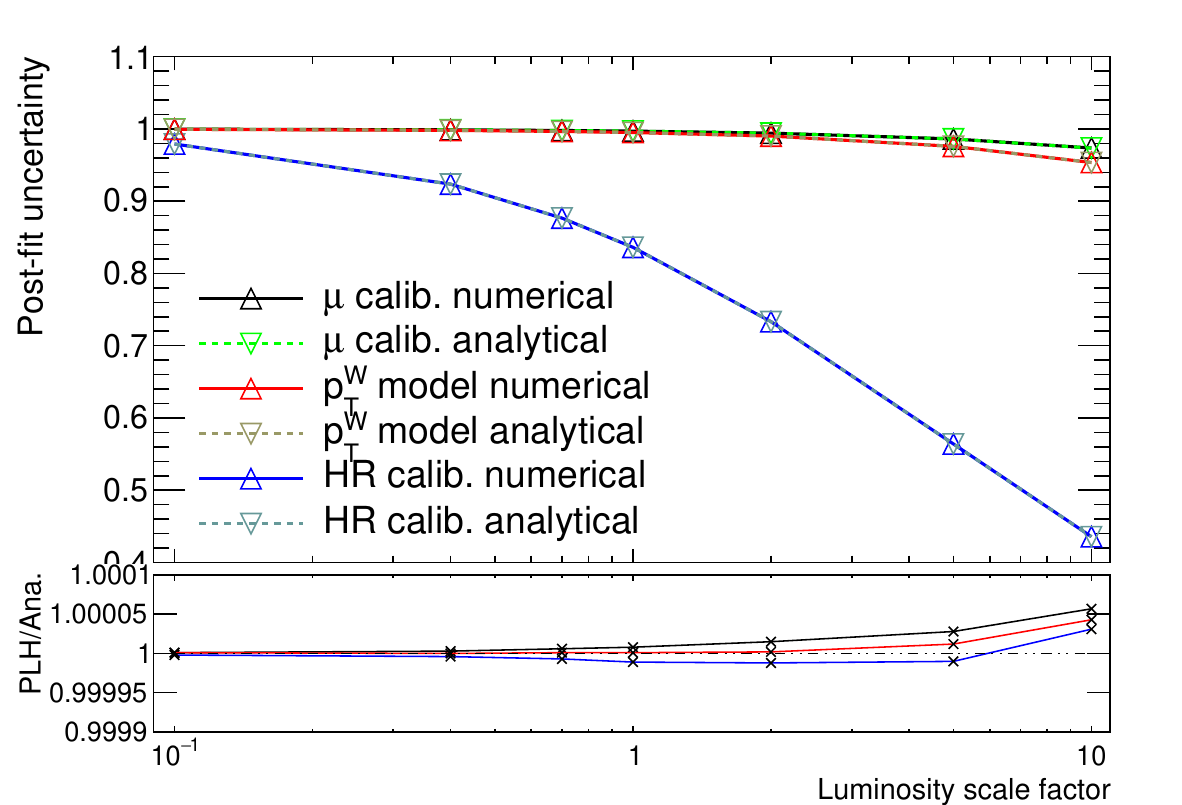}}
\caption{Post-fit NP uncertainties at different values of the luminosity scale factor. The results of the numerical and the analytical PL-fits are compared in the ratio panel.}
\label{fig:fitresult_NPconstraints}
\end{figure*}

Table \ref{tab:mT_ptl_syst_dec_analytical} shows the analytical systematic uncertainty decomposition for the $m_\text{T}$ and $p_\text{T}^\ell$ fits with nominal luminosity, 
together with the NP-POI covariance matrix elements obtained from the numerical profile-likelihood fit. This confirms that the systematic uncertainty components can be directly read from the PL fit covariance matrix, as discussed around Eq.~(\ref{eq:cov_nps_poi_2}). Finally, Figure~\ref{fig:fitresult_NPconstraints} compares the post-fit NP uncertainties between the numerical profile likelihood fit and the analytical calculation. The two methods agree at the 0.1 per-mil level.

\begin{table*}
\begin{adjustbox}{width=1\textwidth}
\begin{tabularx}{\linewidth}{XXX}

\vspace{-1mm}
\begin{tabular}{lcc}\hline
Uncertainty &  $m_\text{T}$ & $p_\text{T}^\ell$ \\ \hline
$\mu$ calib. & 40.67 & 39.65 \\
HR calib. & 12.15 & 0.39 \\
$p_T^W$ model & 2.49 & 9.77 \\ \hdashline
Total & 55.06 & 60.50 \\\hline
\end{tabular}
&
\[
\begin{pmatrix}[ccc:c]
        0.99 & -0.01 & 0.00 & -40.67 \\
        -0.01 & 0.70 & 0.00 & -12.15 \\
        0.00 & 0.00 & 0.99 & -2.49 \\\hdashline
        -40.67 & -12.15 & -2.49 & 3031.64 \\
\end{pmatrix}
\]
    &
\[
\begin{pmatrix}[ccc:c]
        0.99 & 0.00 & -0.03 & -39.65 \\
        0.00 & 1.00 & 0.01 & -0.39 \\
        -0.03 & 0.01 & 0.74 & -9.77\\\hdashline
        -39.65 & -0.39 & -9.77 & 3660.35 \\
\end{pmatrix}
\]
\end{tabularx}
\end{adjustbox}
\caption{Left : list of systematic uncertainty contributions and the total uncertainty, in MeV, for the $m_\text{T}$ and $p_\text{T}^\ell$ fits performed in covariance representation. Centre, right: post-fit covariance among the three NPs associated to these systematic uncertainties and the POI, for the profile-likelihood fits to the $m_\text{T}$ and $p_\text{T}^\ell$ distribution, respectively. \label{tab:mT_ptl_syst_dec_analytical}}
\end{table*}

\subsection{Use of decomposed uncertainties in subsequent fits or combinations}

Uncertainty decompositions obtained with the present method are meaningful only if the results can be used consistently in downstream applications, such as measurement combinations or interpretation fits in terms of specific physics models. In particular, uncertainty components that are common to several measurements generate correlations which should be evaluated properly. This happens when measurements are statistically correlated or when they are impacted by shared systematic uncertainties.

As a final validation of the presented method, we test the combination of profile-likelihood fits of the same observable. Such a combination can be performed either using the decomposed uncertainties, or in terms of the  PL fit outputs, i.e. the fitted values of the POIs and NPs and their covariance matrix. 

The combination is performed starting from Eq.~(\ref{eq:chi2cov}) which, as noted in Section~\ref{sec:unccov}, can be applied to linear measurement averaging by adapting the definition of $t(\vec{\theta})$. In case of a single combined parameter, $t_i=\theta$; for a simultaneous combination of several parameters, $t_i = \sum_{p}{U}_{ip}\theta_p$ where ${U}_{ip}$ is 1 when measurement $i$ is an estimator of POI $p$, and 0 otherwise~\cite{Valassi:2003mu}. This gives :

\begin{equation}
    \begin{split}
        -2\ln\mathcal{L}_\text{cmb}(\vec{\theta}) = \sum_{i,j} &\left( m_i - \sum_{p}{U}_{ip}\theta_p \right) C_{ij}^{-1} \\&
        \times\left(m_j - \sum_{p}{U}_{jp}\theta_p \right),
    \end{split}
    \label{eq:chi2cmb}
\end{equation}

which can be solved as in Section~\ref{sec:unccov}.

As an illustration, we use the $m_W$ fits using the $p_\text{T}^\ell$ and $m_\text{T}$ distributions described in the previous section. In the case of a combination based on the uncertainty decomposition, there are two measurements (the POIs of the $p_\text{T}^\ell$ and $m_\text{T}$ fits), one combined value, and the covariance $C$ is a $2\times 2$ matrix constructed from the decomposed uncertainties using Eq.~(\ref{eq:covdef}).

For a combination based on the PL fit outputs, there are in this example eight measurements (one POI and three NPs in the $p_\text{T}^\ell$ and $m_\text{T}$ fits), four combined parameters, and $C$ is an $8\times 8$ matrix. The diagonal $4\times 4$ blocks are the post-fit covariance matrices of each fit ($p_\text{T}^\ell$ and $m_\text{T}$). The off-diagonal blocks reflect systematic and/or statistical correlations between the $p_\text{T}^\ell$ and $m_\text{T}$ fits, and can be obtained analytically following the methods of Section~\ref{sec:UncDecomp_ShiftedObs}. For two fits $f_1$ and $f_2$ the covariance matrix elements are
\begin{eqnarray}
\small
    \begin{aligned}
        \text{cov}\left(\theta_p^{f_1}, \theta_q^{f_2}\right) &= \sum_k\Delta\theta^{[m_k], f_1}_p\,\Delta
        \theta^{[m_k], f_2}_q + \sum_t\Delta\theta^{[a_t], f_1}_p \Delta\theta^{[a_t], f_2}_q\\
        \text{cov}\left(\alpha_r^{f_1}, \alpha_s^{f_2}\right) &= \sum_k\Delta\alpha^{[m_k],f_1}_r \Delta\alpha^{[m_k],f_2}_s + \sum_t\Delta\alpha^{[a_t], f_1}_r
        \Delta\alpha^{[a_t], f_2}_s \\
        \text{cov}\left(\alpha_r^{f_1}, \theta_p^{f_2}\right) &= \sum_k\Delta\alpha^{[m_k], f_1}_r\Delta\theta^{[m_k], f_2}_p + \sum_t\Delta\alpha^{[a_t], f_1}_r
        \Delta\theta^{[a_t], f_2}_p\\
        \text{cov}\left(\theta_p^{f_1}, \alpha_r^{f_2}\right)  &= \sum_k\Delta\theta^{[m_k], f_1}_p \Delta\alpha^{[m_k], f_2}_r  + \sum_t\Delta\theta^{[a_t], f1}_p \Delta\alpha^{[a_t], f_2}_r
    \end{aligned}\label{eq:analytical_covariance}
\end{eqnarray}
For each matrix element, the first sum is statistical and typically occurs when the fitted distributions are projections of the same data, as is the case for the $p_\text{T}^\ell$ and $m_\text{T}$ distributions in $m_W$ fits. The second sum represents shared systematic sources of uncertainty.

\begin{figure}[htbp]
    \centering
    \includegraphics[width=.5\textwidth]{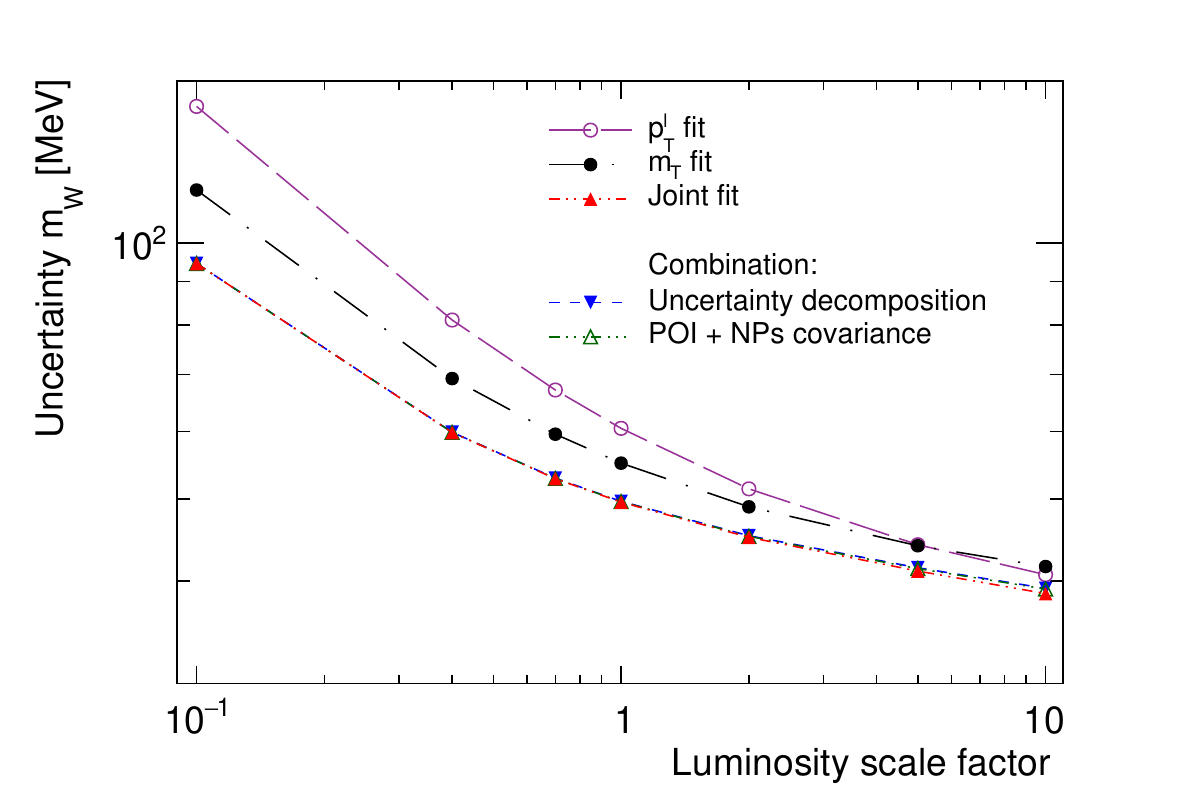}
    \caption{Summary of $m_T$ and $p_T^\ell$ PL fit results. Combinations are produced using the uncertainty decomposition method, and using the covariance of the PL fit results.}
    \label{fig:mw_combination}
\end{figure}

Results of this comparison are presented in Figure~\ref{fig:mw_combination} and Table~\ref{tab:addlabel}, which summarize the fit precision as a function of the assumed luminosity. The uncertainty decomposition method and the combination of the PL fit results agree to better than 0.1~MeV. For completeness, the result of a direct joint fit to the two distributions is shown as well; slightly more precise results are obtained in this case, as expected, especially for high integrated luminosities where systematic uncertainties dominate. 

\begin{table}[htbp]
  \centering
    \begin{tabular}{c|ccc|cc}
    \hline    
    \multicolumn{1}{c|}{} & \multicolumn{5}{c}{Total uncertainty in $m_W$ [MeV]} \\
    \hline
    Luminosity & \multicolumn{3}{c|}{PL fits} & \multicolumn{2}{c}{Combinations} \\
    scale   & \multirow{2}{*}{$m_T$} & \multirow{2}{*}{$p_T^\ell$} & \multirow{2}{*}{Joint} & Unc.  & POI+NP   \\
    factor & & & & decomp. & covariance\\
    \hline
    0.1   & 115.4 & 144.7 & 94.5 & 94.5 & 94.5 \\
    0.4   &  69.2 &  81.1 & 59.8 & 59.8 & 59.9 \\
    0.7   &  59.6 &  67.1 & 52.8 & 52.8 & 52.9 \\
    1.0   &  55.1 &  60.5 & 49.5 & 49.6 & 49.6 \\
    2.0   &  48.9 &  51.4 & 45.0 & 45.2 & 45.2 \\
    5.0   &  44.0 &  44.1 & 41.1 & 41.5 & 41.4 \\
    10.0  &  41.6 &  40.7 & 38.7 & 39.2 & 39.2 \\
    \hline
    \end{tabular}%
  \caption{Summary of $m_T$ and $p_T^\ell$ PL fit results. Combinations are produced using the uncertainty decomposition method, and using the covariance of the PL fit results. \label{tab:addlabel}}
\end{table}%

We note that a combination of PL fit results based on the nuisance parameter representation, Eq.~(\ref{eq:chi2np}), as proposed in Ref.~\cite{Kieseler:2017kxl}, seems difficult to justify rigorously. The principal reason is that Eq.~(\ref{eq:chi2np}) explicitly relies on the absence of correlations, prior to the combination, between the sources of uncertainty encoded in the covariance matrix $V$ and the uncertainties treated as nuisance parameters. Since the input measurements result from PL fits, the POI of each input measurement is in general correlated with the corresponding NPs. One possibility would be to add terms to Eq.~(\ref{eq:chi2np}) that describe these missing correlations. It could also be envisaged to diagonalize the covariance of the inputs and perform the fit in this new basis, but this would work only if all measurements can be diagonalized by the same linear transformation, which is in general not the case.

\section{Conclusion}
\label{sec:conclusion}

We have studied the decomposition of fit uncertainties in two often-used statistical methods in high-energy physics, namely fits in covariance representation and the profile likelihood. The equivalence between the two methods in the Gaussian limit is reminded, and a complete set of expressions is given for the fit uncertainties in the parameters of interest, the nuisance parameters and their correlations. A direct correspondence is established between the standard uncertainty decomposition in covariance representation and the (POI,NP) covariance matrix elements in nuisance representation.

Numerical profile-likelihood analyses generally define statistical and systematic uncertainty components from the results of statistical-only fits and systematic impacts, but this identification does not hold. The uncertainty of statistical-only fits underestimates the statistical uncertainty of fits including systematics, and systematic impacts correspondingly overestimate the genuine systematic uncertainty contributions. Impacts cannot be used as inputs to subsequent measurement combinations or interpretation fits. 

We introduce a set of analytical and numerical methods to remove this shortcoming. In Gaussian approximation, a consistent uncertainty decomposition can be directly extracted from the PL fit covariance matrix. For general (non-gaussian or non-linear) profile-likelihood fits, we propose a procedure, using shifted observables, from which a consistent uncertainty decomposition can be obtained rigourously. We illustrate these points by means of simple examples, and show that profile-likelihood fit results with properly decomposed uncertainties can be used consistently in downstream combinations or fits.


\bibliographystyle{spphys}
\bibliography{EPJC}

\end{document}